\documentclass{elsart}
\usepackage{graphicx}
\usepackage{epsfig}
\usepackage{amssymb}

\renewcommand{\baselinestretch}{1.25}

\journal{Physics Letters B}

\begin{document}
%-----------------------------------------------------------------------
\begin{frontmatter}

\title{Applicability of Perturbative QCD \\
 to Pion Virtual Compton Scattering}

\author[pku]{Jian Su},
\author[ccastpku]{Bo-Qiang Ma\corauthref{cor}}
\corauth[cor]{Corresponding author.} \ead{mabq@phy.pku.edu.cn}
\address[pku]{Department of Physics, Peking University, Beijing 100871, China}
\address[ccastpku]{CCAST (World Laboratory), P.O.~Box 8730, Beijing
100080, China\\
Department of Physics, Peking University, Beijing 100871, China}

\begin{abstract}
We study explicitly the applicability of perturbative QCD (pQCD)
to the pion virtual Compton scattering. It is found that there are
central-region singularities introduced by the QCD running
coupling constant, in addition to the end-point singularities
generally existed in other exclusive processes such as the pion
form factor. We introduce a simple technique to evaluate the
contributions from these singularities, so that we can arrive at a
judgement that these contributions will be unharmful to the
applicability of pQCD at certain energy scale, i.e., the ``work
point'' which is defined to determine when pQCD is applicable to
exclusive processes. The applicability of pQCD for different pion
distribution amplitudes are explored in detail. We show that pQCD
begins to work at 10 $\mbox{GeV}^2$. If we relax our constraint to
a weak sense, the work point may be as low as 4 $\mbox{GeV}^2$.
\end{abstract}


\end{frontmatter}

%-----------------------------------------------------------------------
%\section{Introduction}
%-----------------------------------------------------------------------

The application of perturbative Quantum Chromodynamics (pQCD) to
exclusive processes started since late 1970's
\cite{BL,Lep80,BHL81}, and now the pQCD approach has been widely
employed to various exclusive processes. It is generally known
that pQCD is applicable at very high energy scale, and one of the
challenging questions is whether pQCD is valid or not at present
experimental accessible energy scale. A typical example is the
applicability of pQCD to the pion form factor. It should be
actually unknown whether pQCD is applicable or not before any
affirmative judgement that the dominance of perturbative effect
over the soft effect is justified. It was indicated by Isgur and
Llewellyn~Smith \cite{IL} that a significant fraction of
contributions to the form factor is from the soft end-point
regions at medium-to-high energy scale. In calculating the pion
electromagnetic form factor, Huang and Shen \cite{HS} showed that
the non-perturbative contribution from end-point regions can be
suppressed with the end-point-suppressed distribution amplitude by
taking into account the transverse momentum in the pion wave
function. Li and Sterman \cite{LS} reduced the end-point
contribution by Sudakov effect, i.e., to replace the end-point
singularity by theoretical based resummation over soft radiative
corrections. There is also an attempt \cite{Yeh} to explain the
large discrepancy between theory and experiment by large
contribution from next-to-leading-order power corrections.

One of the alternative exclusive processes, Compton scattering,
which is another good laboratory to study the structure of
hadrons, has been extended from real photon to virtual photon for
the proton case as well as the pion case. Several independent
theory groups have calculated the pion virtual Compton scattering
(VCS) with very different methods, such as the chiral perturbation
theory \cite{U}, the QCD sum rules \cite{CL}, and pQCD approach
\cite{CL,T,MT,ZM}. One should meet the same question concerning
the validity of pQCD, in similar to the case of the pion form
factor. It is the purpose of this paper to study explicitly the
applicability of pQCD to the pion VCS process. The early
literature focused on how to eliminate the specific problem of
kinematic singularities by analytic integration, but lack of a
full study of the end-point effect.  We will show that there are
central-region singularities in the pion VCS, in additional to the
end-point singularities generally existed in other exclusive
processes such as the pion form factor. We will introduce a
technique to evaluate the contributions from these singularities,
so that we can have a criterion for the applicability of pQCD at
certain energy scale.

%\section{Kinematics and Formulae}

\begin{figure}[tbh]
\begin{center}
\scalebox{0.5}[0.5]{\includegraphics*[53pt,246pt][491pt,578pt]{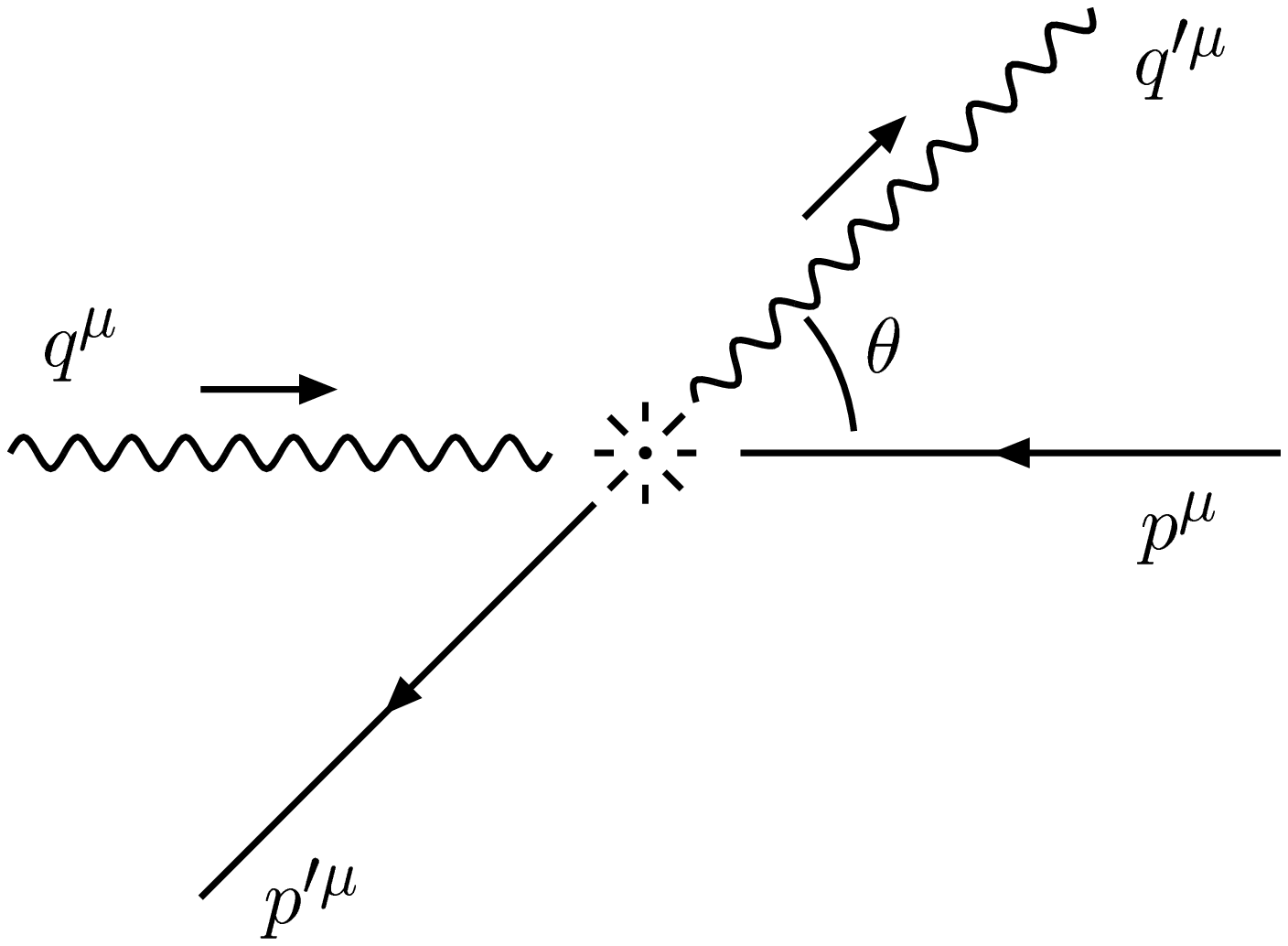}}
\end{center}
\caption{\footnotesize Virtual Compton scattering $\pi\gamma^*\to
\pi\gamma$ shown in the center-of-mass frame.} \label{CM}
\end{figure}

The pion VCS process is illustrated in Fig.~\ref{CM}, where
$p^{\mu}$ and $q^{\mu}$ are the 4-momenta of the incoming pion and
virtual photon respectively, and ${p'}^{\mu}$ and ${q'}^{\mu}$ are
the corresponding 4-momenta of the outgoing pion and photon. We
define the following frame invariant variables:
\begin{equation}
\begin{array}{ll}
Q^2=-q^2, \ \ \ \ S=(q+p)^2=({q'}+{p'})^2, \\
t=({p'}-p)^2, \ \ \ \ u=({p'}-q)^2,
\end{array}
\end{equation}
where $q^2$ is the squared 4-momentum of the virtual photon, and
$S$ is the squared center-of-mass energy of the virtual photon and
pion system.

%\vspace{2.0cm}

Factorization theorem \cite{BL,Lep80} tells us that the scattering
amplitude of an exclusive process can be expressed in a
convolution formula:
\begin{eqnarray}
M(Q^2)=\int_0^1[dx]\int_0^1[dy]
\phi^*(y_i,\tilde{Q}_y)H(x_i,y_i,\tilde{Q}^2)\phi(x_i,\tilde{Q}_x)[1+O(m^2/Q^2)],
\label{FT}
\end{eqnarray}
where in the pion VCS case $[dx]\equiv
dxd\bar{x}\delta(1-x-\bar{x})$, $\bar{x}\equiv1-x$,
$\tilde{Q}_x\equiv\min(xQ,\bar{x}Q)$, $x$ is the light-cone
$k^+=k^0+k^3$ momentum fraction of the struck quark (antiquark) in
the incoming pion, and $y$ is the light-cone momentum fraction of
the corresponding quark (antiquark) in the outgoing pion. The
validity of Eq.~(\ref{FT}) is that there is an initial $Q=Q_0$ as
the factorization scale above which the hard part and the soft
part could be separated safely and explicitly. We expect $S$,
$-t$, and $-u$ large to guarantee factorization legal in the VCS
case. The distribution amplitude $\phi(x,Q)$ may be chosen without
higher Gegenbauer polynomials to exclude $Q$ dependence. In this
paper the asymptotic (as) \cite{BL,Lep80} and Chernyak-Zhitnitsky
(cz) \cite{CZ} distribution amplitudes:
\begin{eqnarray}
\phi_{as}&=&\sqrt{3}f_\pi x(1-x),\\
\phi_{cz}&=&5\sqrt{3}f_\pi x(1-x)(1-2x)^2,
\end{eqnarray}
and their end-point-suppressed partners of Brodsky-Huang-Lepage
(bhl) \cite{BHL81,Huang94} and Huang-Shen (hs) \cite{HS}
distribution amplitudes:
\begin{eqnarray}
\phi_{bhl}&=&1.4706\sqrt{3}f_\pi
x(1-x)\exp[\frac{-0.07043}{x(1-x)}],\\
\phi_{hs}&=&8.8763\sqrt{3}f_\pi x(1-x)(1-2x)^2 \exp[
\frac{-0.07062}{x(1-x)}],
\end{eqnarray}
are used and compared in our calculations, where the pion decay
constant $f_\pi=93$~MeV is adopted. We use the leading-twist
factorization scheme instead of the handbag scheme. The different
hard scattering amplitudes $H$ have the following relations due to
parity symmetry \cite{ZM}:
\begin{equation}
\begin{array}{ll}
H_{LL}=H_{RR},\ \ H_{RL}=H_{LR},  \\
H_{+L}=H_{+R},\ \ H_{-R}=-v^{-1}H_{+R}, \ \ H_{_L}=-v^{-1}H_{+L},
\end{array}
\nonumber
\end{equation}
thus only 3 independent $H_{LR}, H_{RR}, H_{+R}$ needed to be
evaluated. The expressions of the hard scattering amplitudes $H'$
\cite{ZM} , which have been multiplied by the prefactor
$x\bar{x}y\bar{y}$ of distribution amplitudes and not included
$\frac{1}{S}\alpha_e\alpha_s$ yet, are listed in Table \ref{tab
1}, where the notations for the diagrams are corresponding to
Fig.~2 of Ref.~\cite{ZM}.

\renewcommand{\baselinestretch}{1.0}
\begin{table*}
\caption{\label{tab 1}The hard scattering amplitudes calculated by
pQCD}
\begin{tabular}{lccc}
\hline\hline\\
diagram & $H'_{LR}$ &$H'_{RR}$ &$H'_{+R}$\\
\hline   \\
a &$\frac{20xy}{9\bar{v}(x-a)}$ &$\frac{20xyc^2}{9\bar{v}s^2
(x-a)}$&$\frac{40cxy}{9s}$\\
&  &   \\
b &$\frac{20x}{9\bar{v}}$ &$\frac{20xc^2}{9\bar{v}s^2}$&$\frac{20cx}{9\bar{v}s}$\\
&  &  \\
c &$0$ &$\frac{-20xc^2}{9\bar{v}s^2(x-a)}$ &$ \frac{-20cx}{9s}$\\
&  & \\
d&$\frac{16x\bar{x}\
\bar{y}s^2}{9c^2}(\frac{1}{x-a}-\frac{1}{x-b})$
&$\frac{16x\bar{y}(1-\bar{v}\bar{x}s^2)}{9\bar{v}c^2}(\frac{1}{x-a}-\frac{1}{x-b})$
&$\frac{-16sxy\bar{y}(1-2\bar{v}\bar{x}s^2)}{9\bar{v}c(1-ys^2)(x-b)}$\\
&  &  \\
e & $ \frac{-16x\bar{x}s^2}{9c^2}(\frac{1}{x-a}-\frac{1}{x-b})$
&$0$&$ \frac{16csx\bar{x}y}{9(1-ys^2)(x-b)}$\\
&  &  \\
f & $0$ &$ \frac{20y}{9\bar{v}s^2}$ &$ \frac{20y(1-2\bar{v}x)}{9\bar{v}cs}$\\
&  &  \\
g & $ \frac{-20axy}{9[y(1-\bar{v}s^2)-v]}$
 &$ \frac{-20xyc^2}{9\bar{v}s^2[y(1-\bar{v}s^2)-v]}$
 &$ \frac{-20caxy}{9s[y(1-\bar{v}s^2)-v]}$\\
&  &  \\
h &$\frac{-20y[y+\bar{v}x(1-2ys^2)]}{9\bar{v}[y(1-\bar{v}s^2)-v]}$
 &$\frac{-20y[1-s^2(y+\bar{v}x)+2\bar{v}xys^4)]}{9\bar{v}s^2[y(1-\bar{v}s^2)-v]}$
 &$\frac{-20cy^2(1-2\bar{v}xs^2)}{9\bar{v}s[y(1-\bar{v}s^2)-v]}$\\
&  &  \\
i&$\frac{-16x\bar{x}c^2s^2y^2}{9(1-ys^2)[y(1-\bar{v}s^2)-v](x-b)}$
 &$\frac{-16x\bar{x}yc^2}{9[y(1-\bar{v}s^2)-v](x-b)}$
 &$ \frac{-16sx\bar{x}y^2c^3}{9(1-ys^2)[y(1-\bar{v}s^2)-v](x-b)}$\\
&  &  \\
j & $
\frac{-16xy\bar{y}s^2[\bar{y}-\bar{v}\bar{x}(1-2ys^2)]}{9\bar{v}(1-ys^2)[y(1-\bar{v}s^2)-v](x-b)}$
 &$ \frac{-16xy\bar{y}s^2[y-v+\bar{v}\bar{x}(1-2ys^2)]}{9\bar{v}(1-ys^2)[y(1-\bar{v}s^2)-v](x-b)}$
 &$ \frac{-16csxy\bar{y}[v-y+2\bar{v}\bar{x}ys^2]}{9\bar{v}(1-ys^2)[y(1-\bar{v}s^2)-v](x-b)}$\\
 \hline\hline\\
\end{tabular}
\end{table*}
\renewcommand{\baselinestretch}{1.25}

%-----------------------------------------------------------------------
%\section{Calculations}
%-----------------------------------------------------------------------

For simplicity, we neglect the mass of pion and work in the
center-of-mass frame (Fig.~\ref{CM}) following available pQCD
studies \cite{CL,T,MT,ZM}. The 4-momenta of the incoming photon
and pion are:
\begin{eqnarray}
& q^{\mu}=(\omega,0,0,p),\ \ p^{\mu}=(p,0,0,-p),\nonumber
\end{eqnarray}
and the corresponding 4-momenta of the outgoing particles are:
\begin{eqnarray}
&{q'}^{\mu}=\frac{\omega+p}{2}(1,\sin \theta,0,\cos \theta),\ \
{p'}^{\mu}=\frac{\omega+p}{2}(1,-\sin \theta,0,-\cos \theta).
\nonumber
\end{eqnarray}
Here we define
\begin{equation}
\begin{array}{ll}
S\equiv(q^{\mu}+p^{\mu})^2=(\omega+p)^2, \\
c\equiv\cos\frac{\theta}{2},\ s\equiv\sin \frac{\theta}{2},\ \
\\
v\equiv q^2/S, \ \ \bar{v}\equiv1-v, \\
a\equiv-\frac{v}{\bar{v}}, \ \
b\equiv\frac{y-v-y\bar{v}s^2}{\bar{v}(1-ys^2)}, \\
t\equiv({p'}^{\mu}-p^{\mu})^2=-S\bar{v}s^2,
\end{array}
\end{equation}
where $v$ stands for the photon virtuality and should be between
$-1$ and $0$, and $\theta$ is the angle between the incoming
virtual photon and the outgoing photon and can be obtained from
the experimental variables by
\begin{equation}
s^2=\sin^2 \frac{\theta}{2}=-\frac{t}{S \bar{v}}=-\frac{t}{S-q^2}.
\end{equation}
We let $v=-0.8$ as an example in our calculations to ensure $q^2$
large. $-t$ is monotonously increasing with the scattering angle
$\theta$ if $S$ and $v$ are both fixed. For diagrams a, b, c, f, g
and h (see Fig.~2 of \cite{ZM}), the squared 4-momentum transfer
of the exchanged gluons between the two valence partons in the
pion is
\begin{equation}
\tilde{Q}^2=-\bar{v}\bar{x}\bar{y}S,
\end{equation}
where the small $\tilde{Q}^2$ region is also the end-point region
whose effects will be highly suppressed by the distribution
amplitudes. For diagrams d, e, i and j, where the incident and
outgoing photons are connected to different quark lines, the
corresponding squared 4-momentum transfer of the exchanged gluons
is
\begin{equation}
\tilde{Q}^2=\bar{v}(1-ys^2)(x-b)S.
\end{equation}
Because of $0<a<b<1$, whatever the kinematical region we will
choose, it is impossible to guarantee $\tilde{Q}^2$ large enough
for making pQCD legal here. The additional singularity of
$\tilde{Q}^2$ in the central region, i.e. $\tilde{Q}^2\to 0$ when
$x\to b$, makes calculations here much more complex than that of
form factor. We use principle integration formula as in \cite{ZM}
\begin{eqnarray}
\lim_{\epsilon \rightarrow 0}\int_0^1
\frac{f(x)}{x-a+i\epsilon}dx&=&P\int_0^1\frac{f(x)}{x-a}dx-i\pi f(a)\nonumber\\
&=&\int_0^1\{\frac{f(x)-f(a)}{x-a}+f(a)[\log\frac{1-a}{a}-i\pi]\}dx
\end{eqnarray}
to get reliable numerical results.

\begin {figure}[h]%2
\scalebox{0.175}[0.175]{\includegraphics*[36pt,0pt][760pt,590pt]{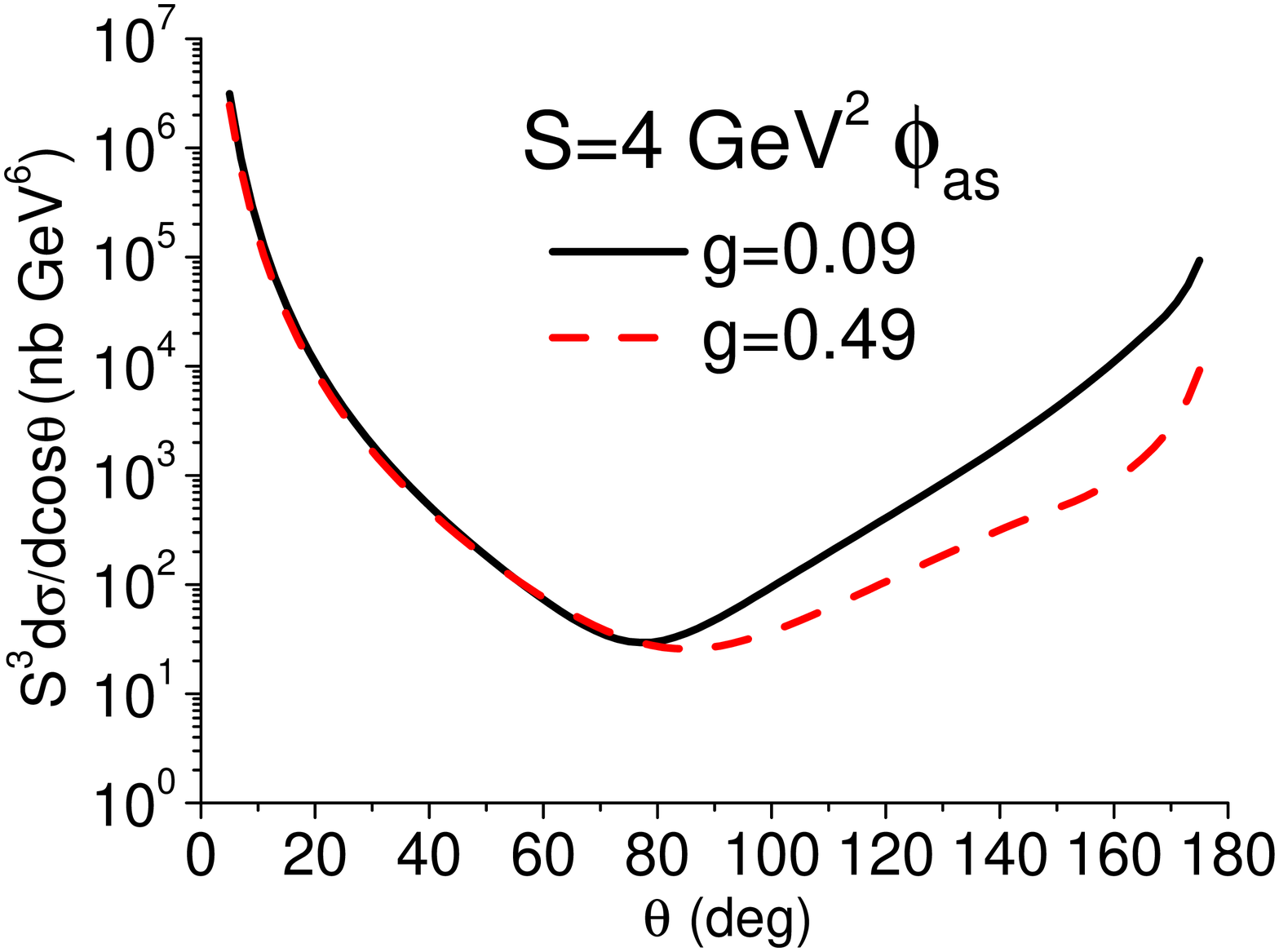}}
\scalebox{0.175}[0.175]{\includegraphics*[36pt,0pt][760pt,590pt]{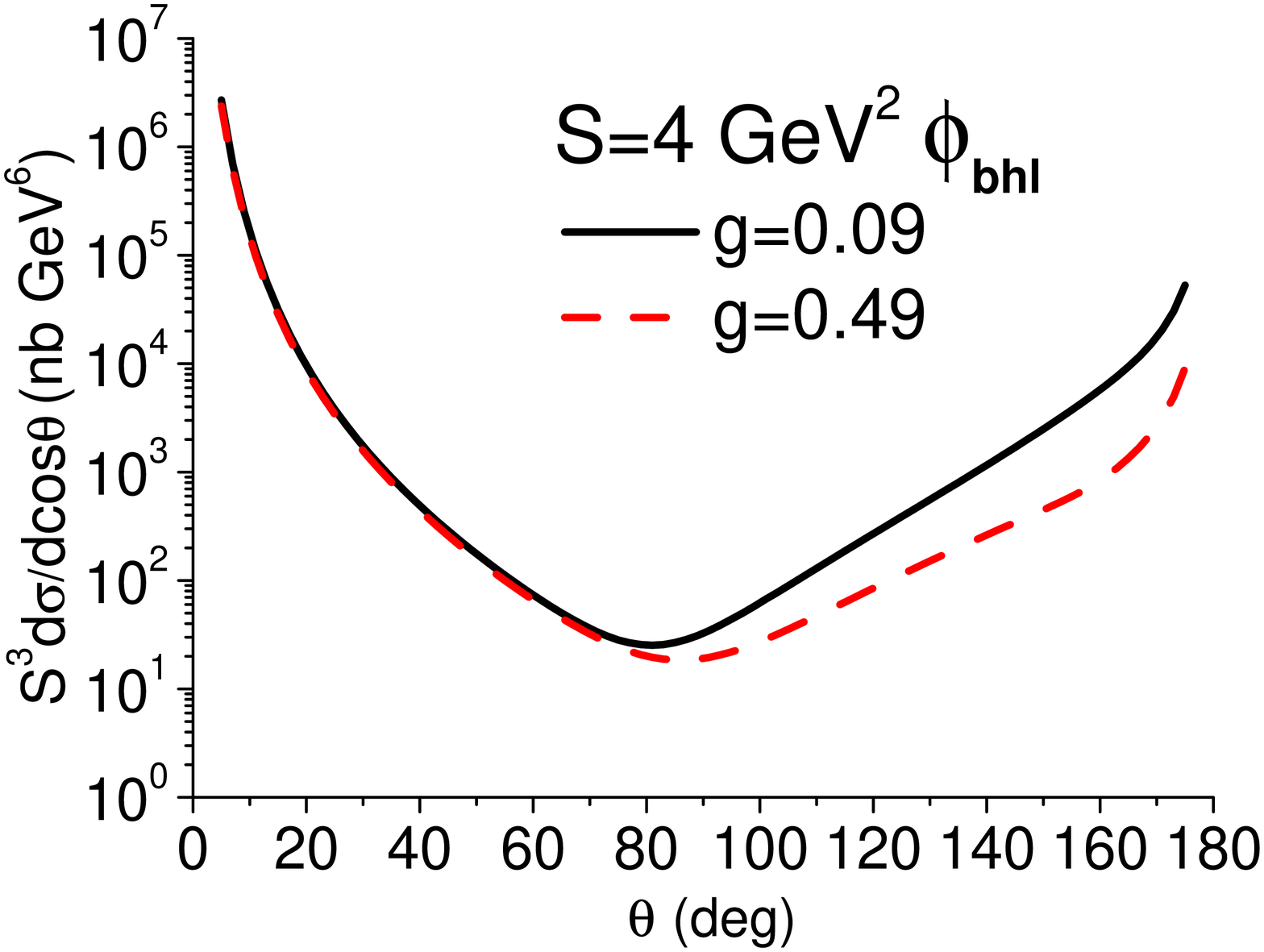}}
\scalebox{0.175}[0.175]{\includegraphics*[36pt,0pt][760pt,590pt]{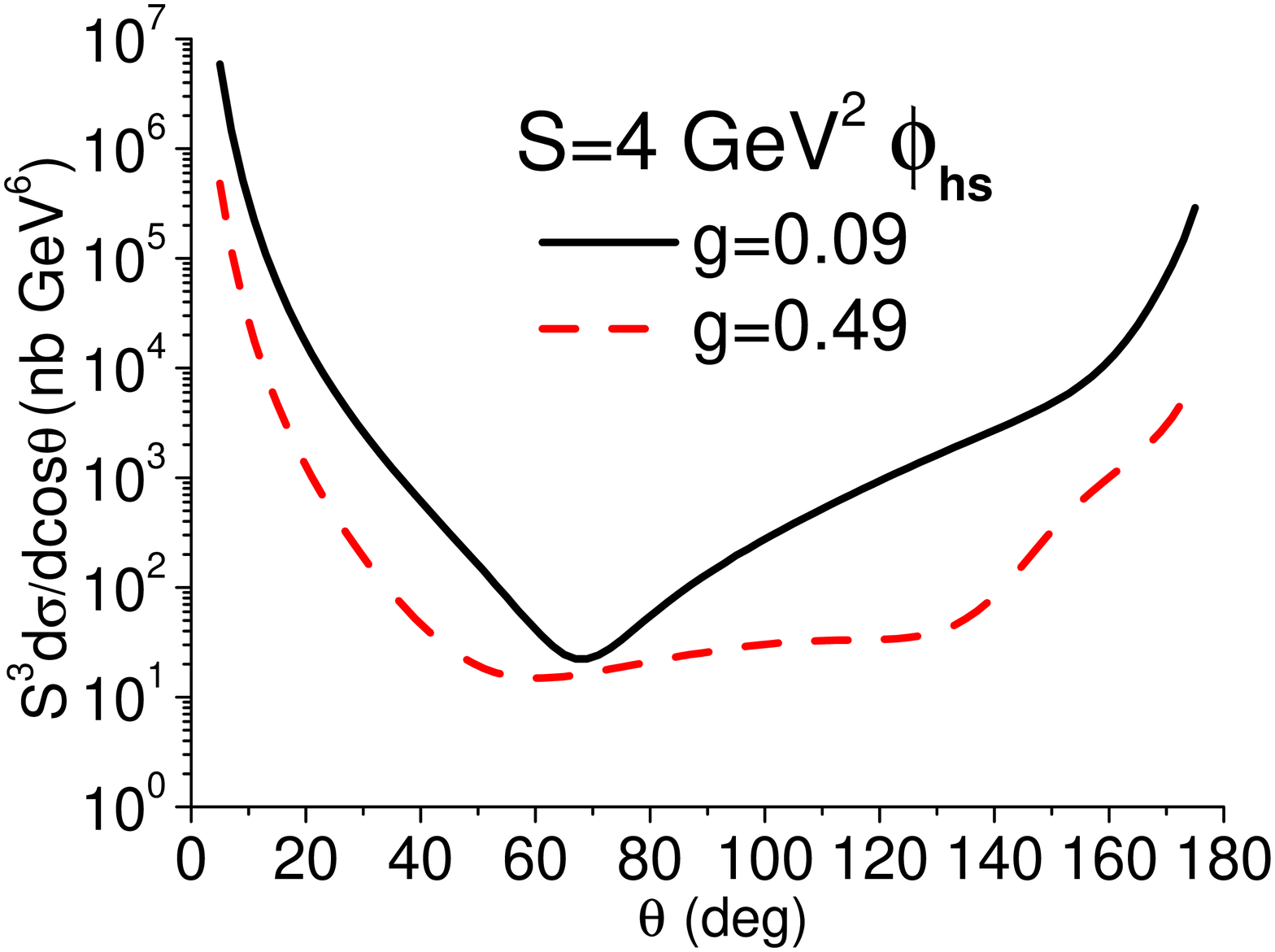}}
\caption{\footnotesize Cross sections at
$S=4~\mbox{GeV}^2$.}\label{4}
\end {figure}

\begin {figure}[h]%3
\epsfxsize=7cm\epsfysize=4.8cm\epsfbox{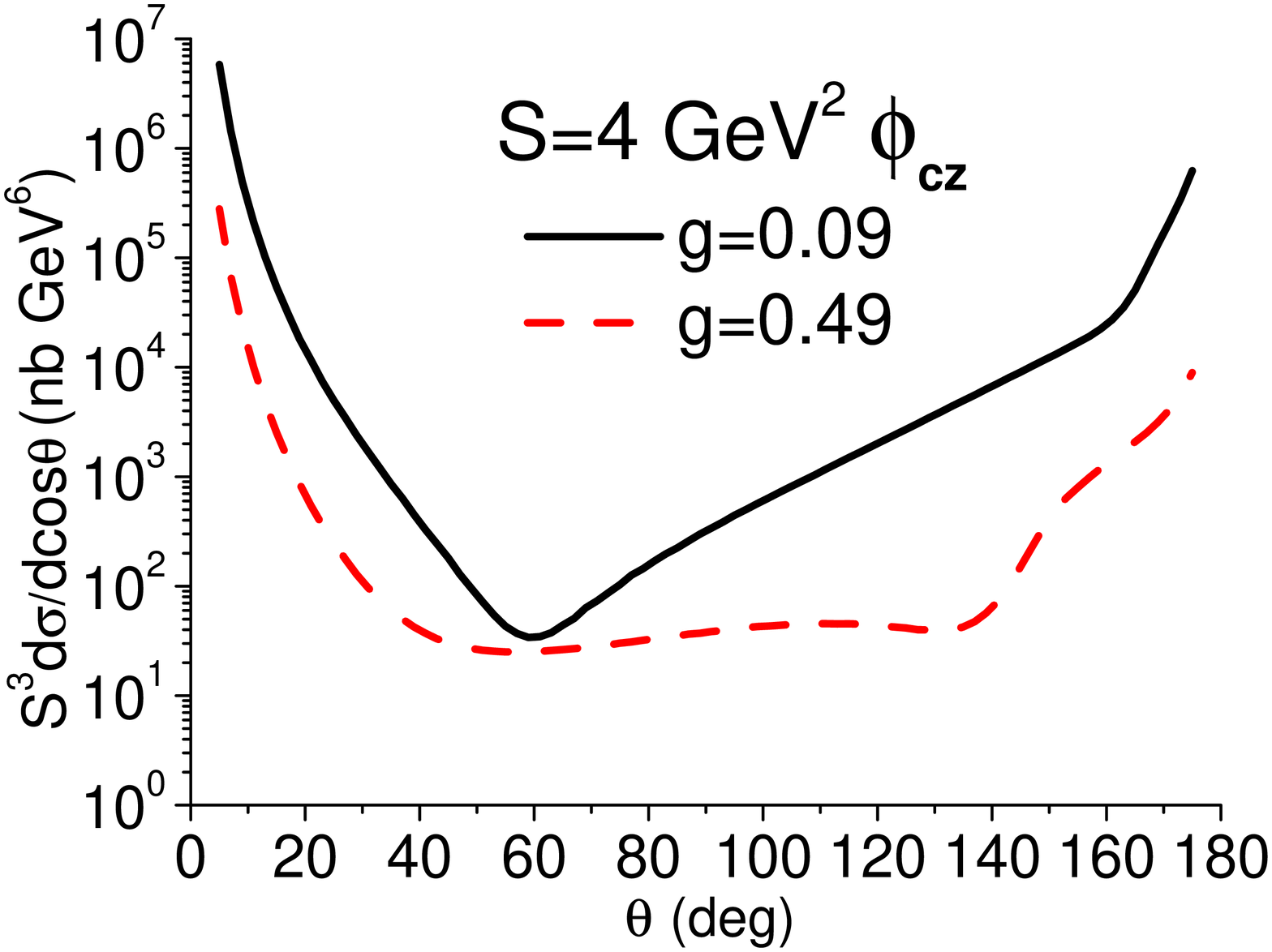}
\epsfxsize=7cm\epsfysize=4.8cm\epsfbox{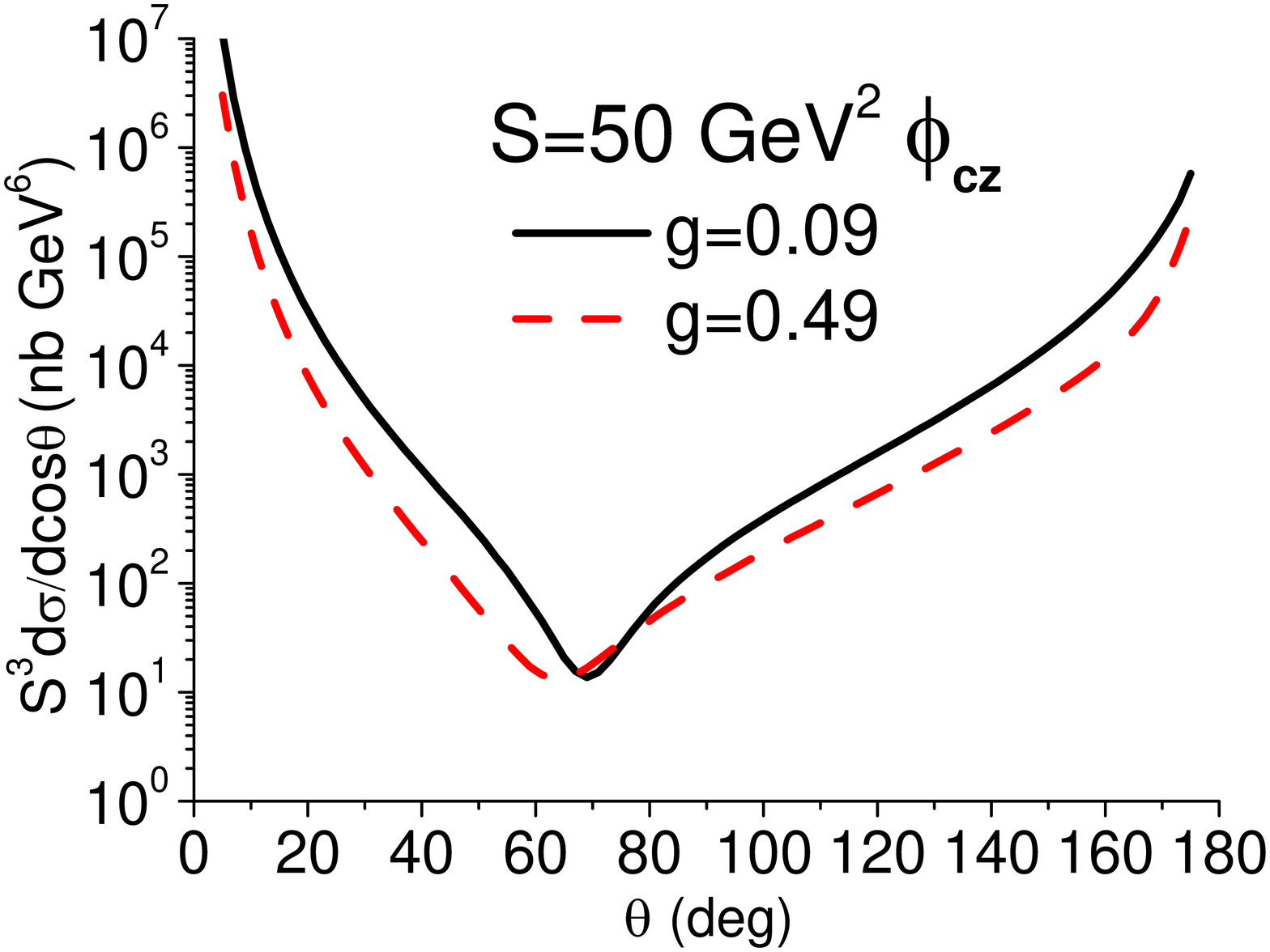}
\caption{\footnotesize Cross sections of $\phi_{cz}$.} \label{cz}
\end {figure}

\begin {figure}[h]%4
\scalebox{0.175}[0.175]{\includegraphics*[36pt,0pt][760pt,590pt]{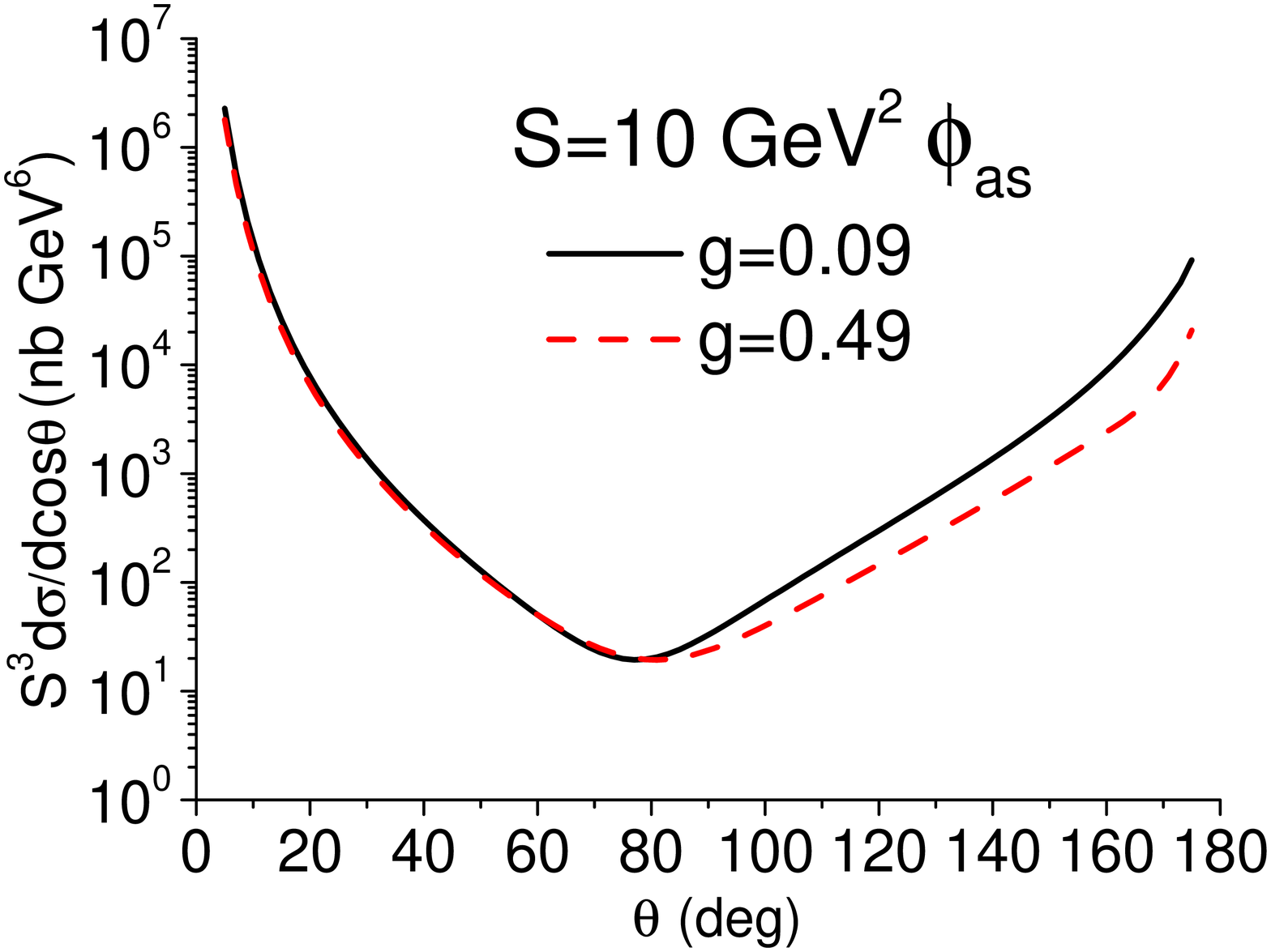}}
\scalebox{0.175}[0.175]{\includegraphics*[36pt,0pt][760pt,590pt]{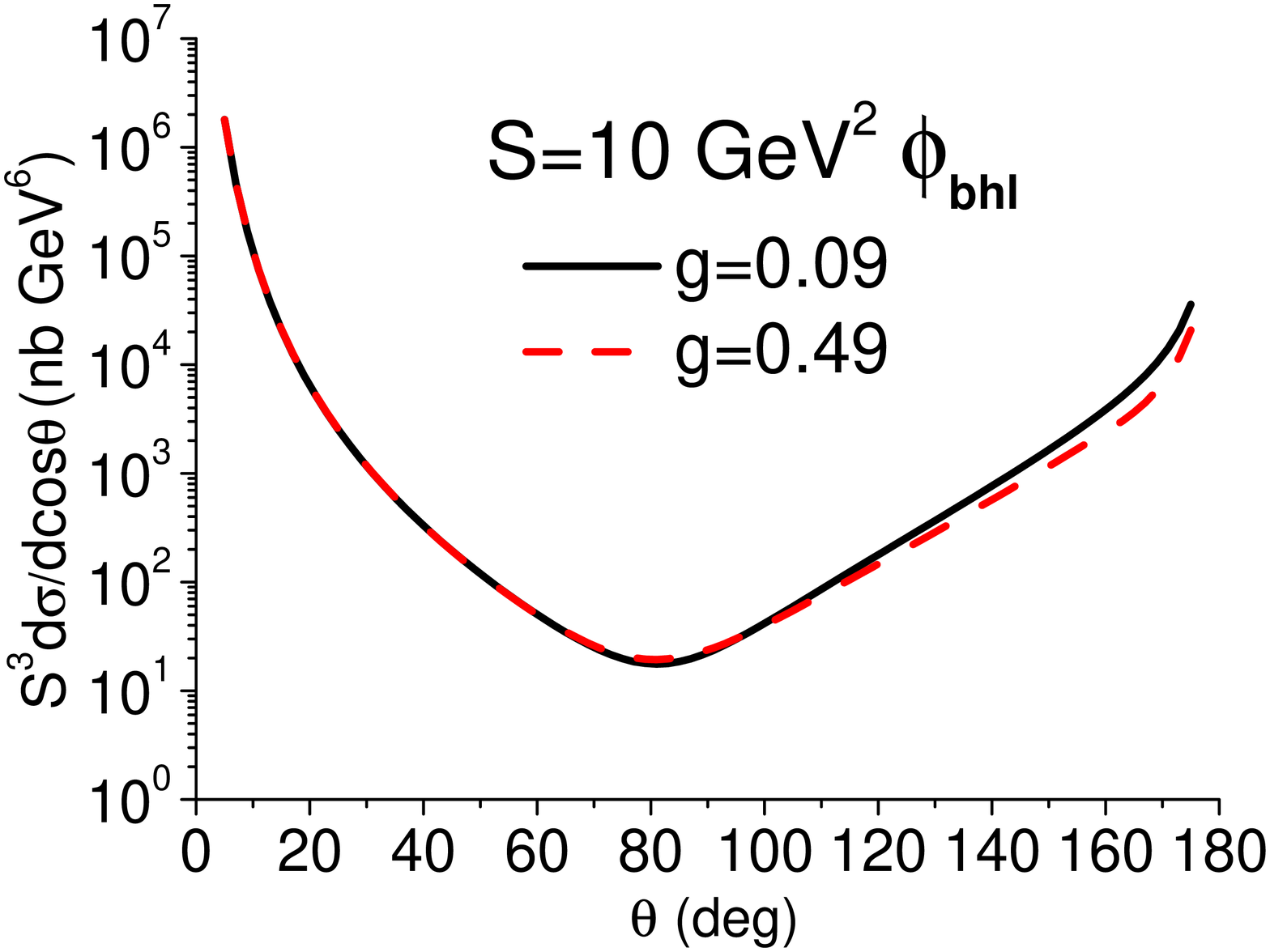}}
\scalebox{0.175}[0.175]{\includegraphics*[36pt,0pt][760pt,590pt]{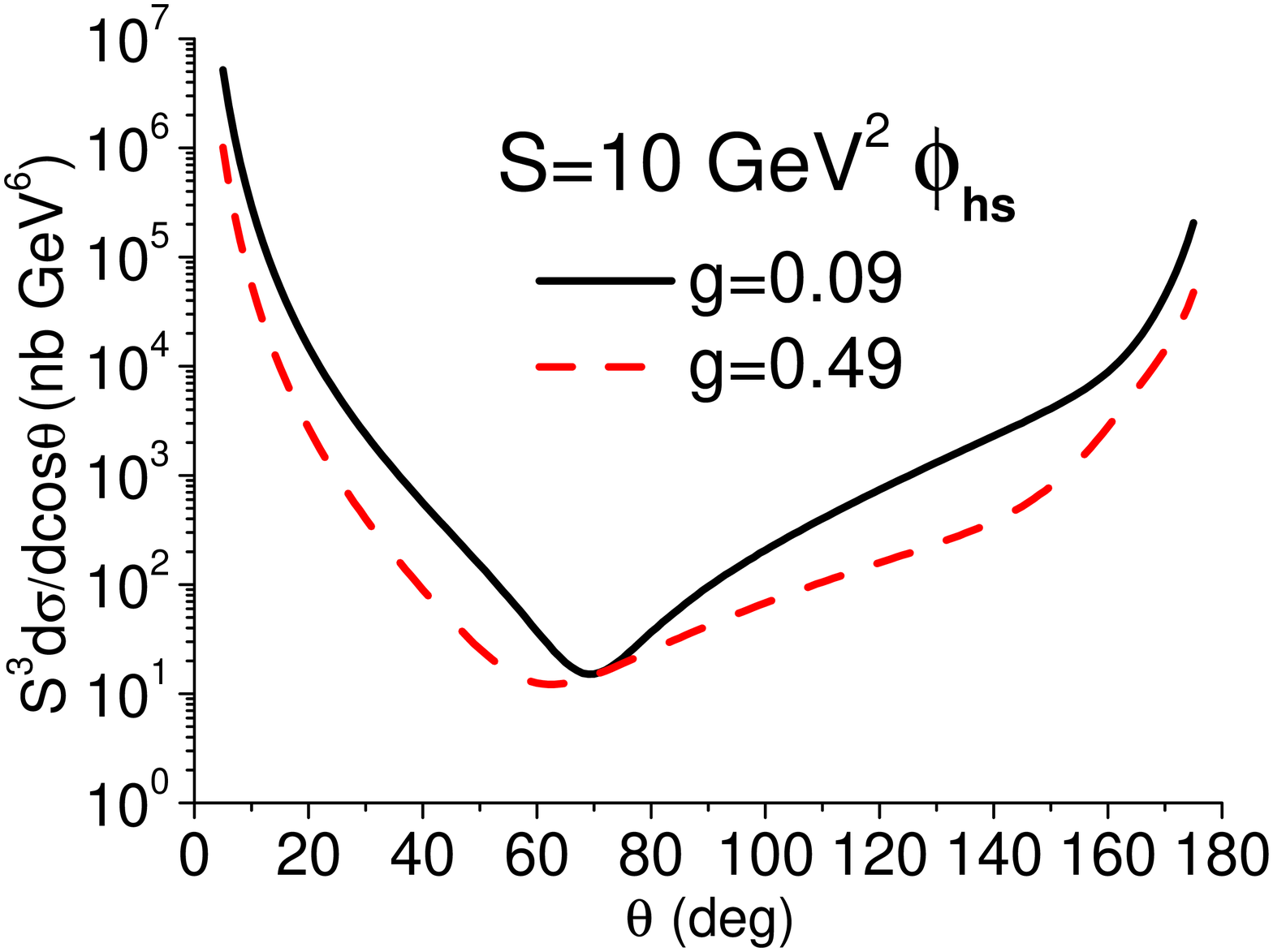}}
\caption{\footnotesize Cross sections at $S=10~\mbox{GeV}^2$.}
\label{10}
\end {figure}

\begin {figure}[h]%5
\scalebox{0.175}[0.175]{\includegraphics*[36pt,0pt][760pt,590pt]{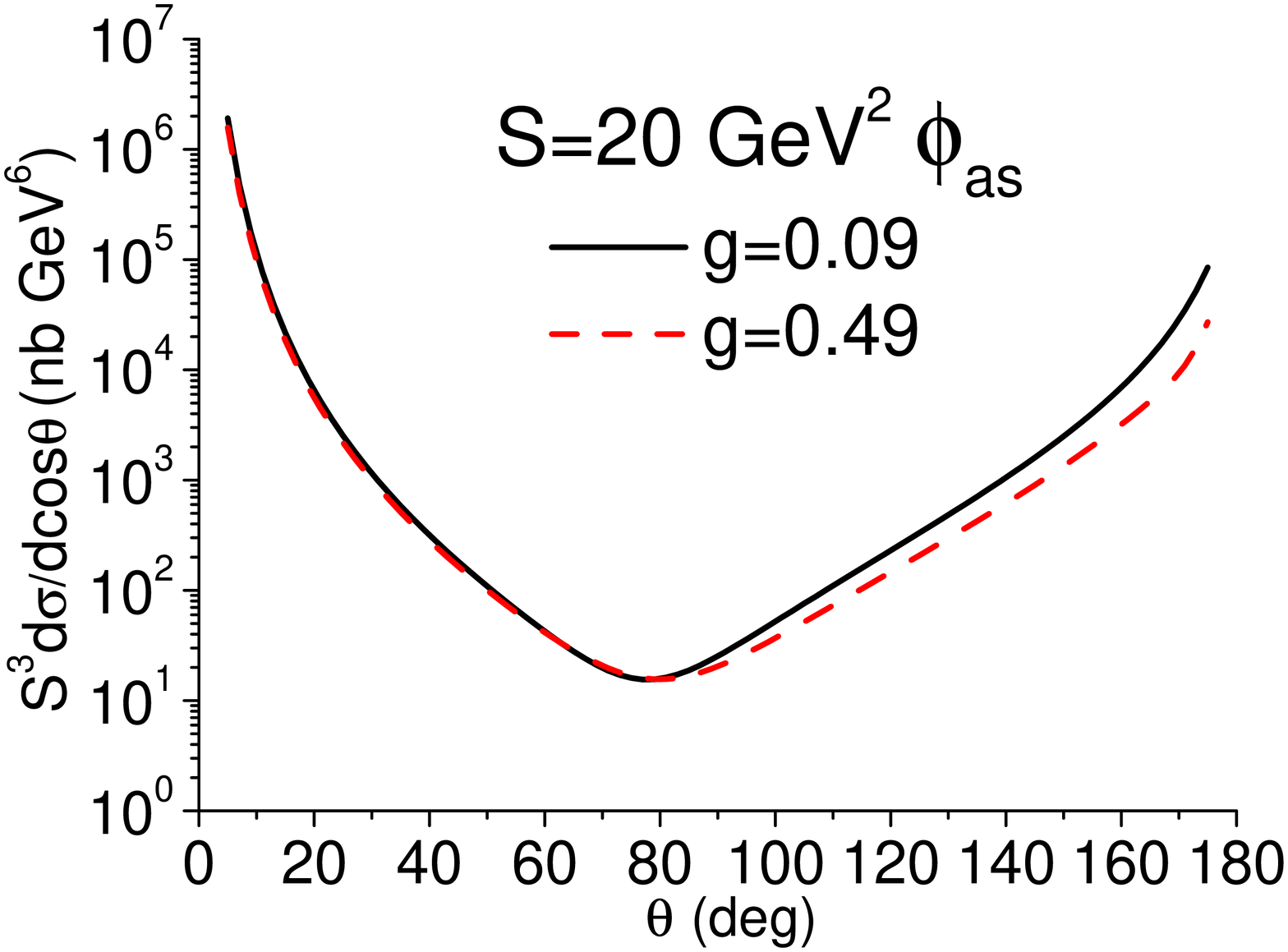}}
\scalebox{0.175}[0.175]{\includegraphics*[36pt,0pt][760pt,590pt]{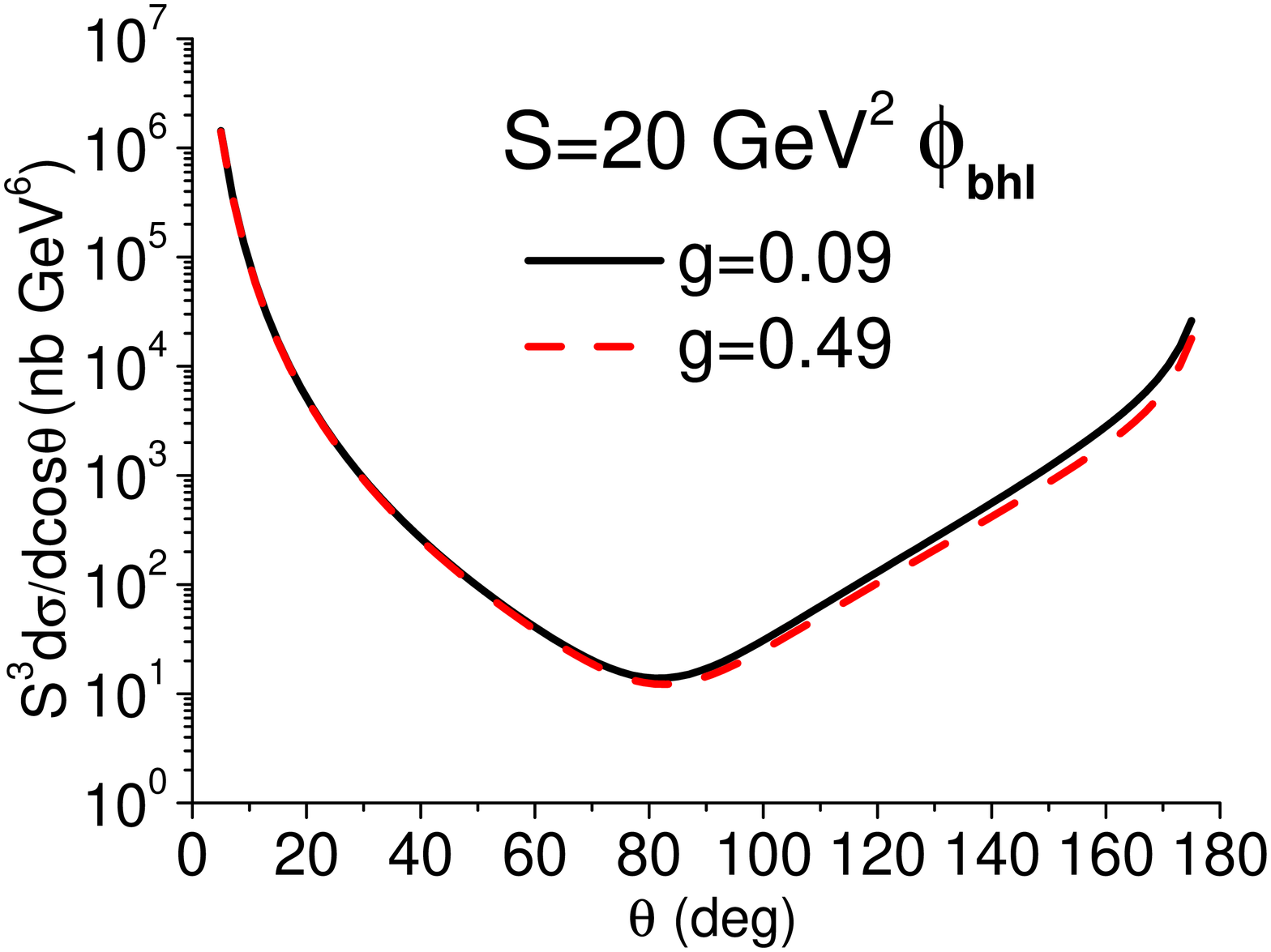}}
\scalebox{0.175}[0.175]{\includegraphics*[36pt,0pt][760pt,590pt]{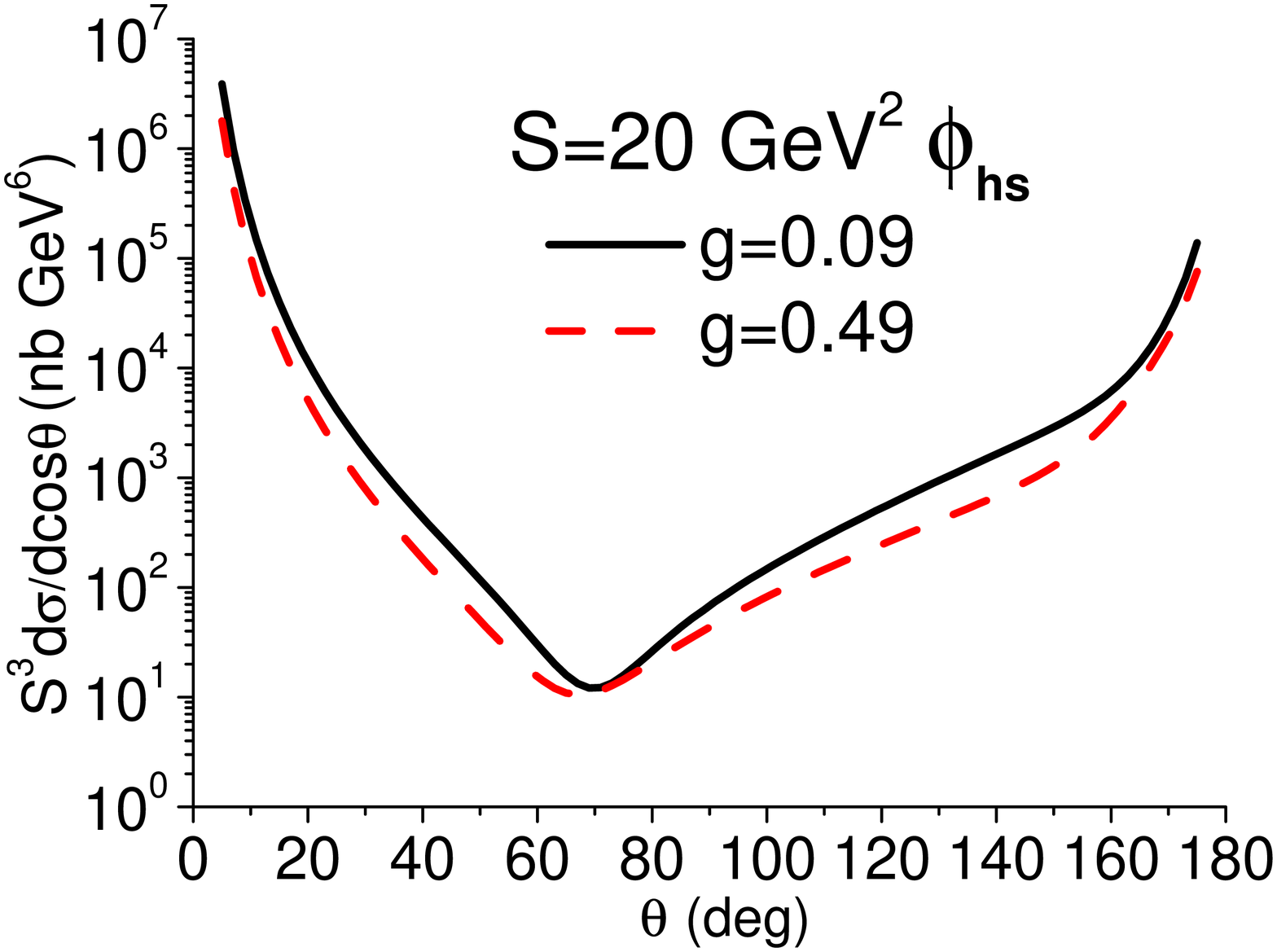}}
\caption{\footnotesize Cross sections at $S=20~\mbox{GeV}^2$.}
\label{20}
\end {figure}

\begin {figure}[h]%6
\scalebox{0.175}[0.175]{\includegraphics*[36pt,0pt][760pt,590pt]{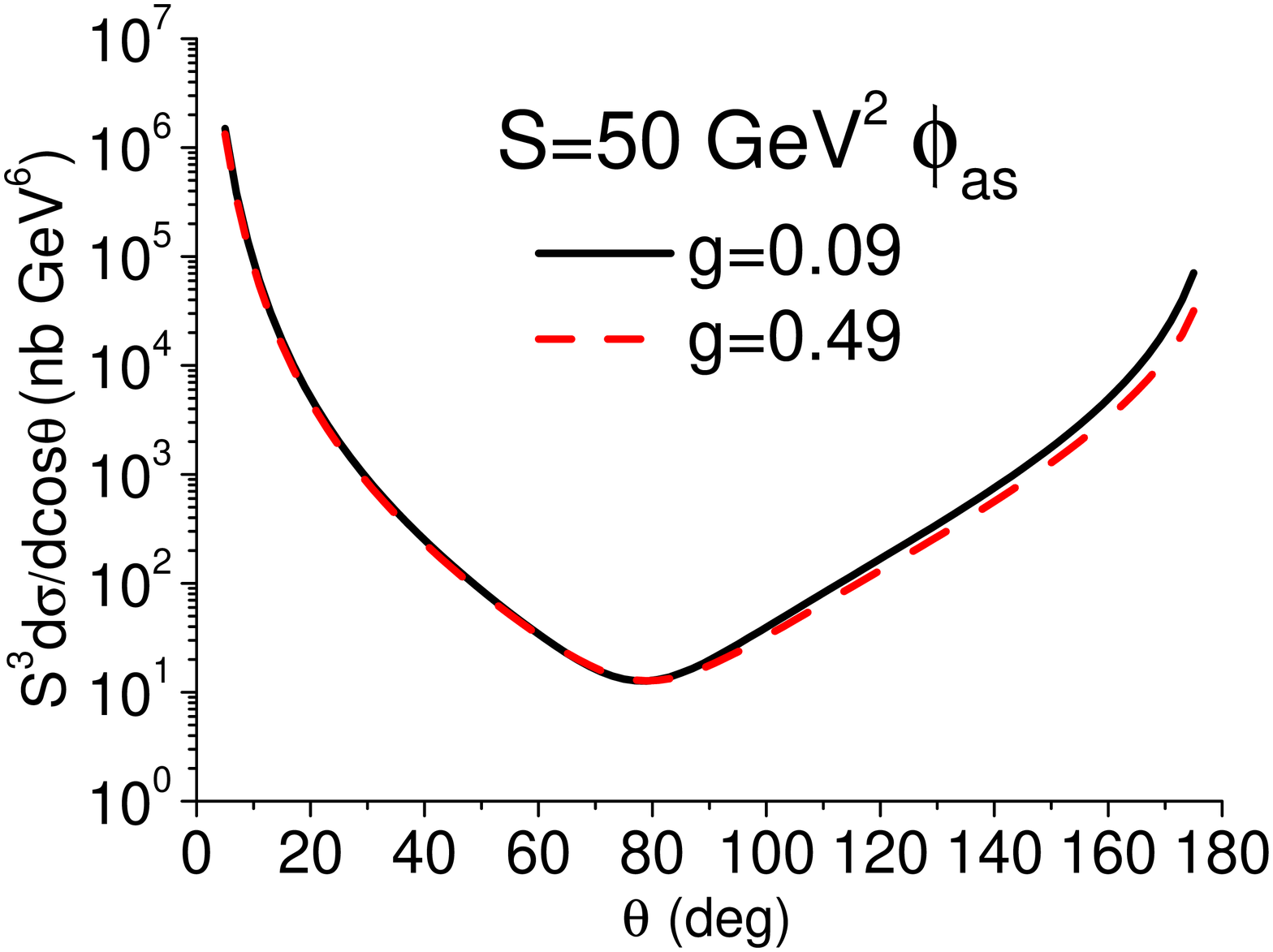}}
\scalebox{0.175}[0.175]{\includegraphics*[36pt,0pt][760pt,590pt]{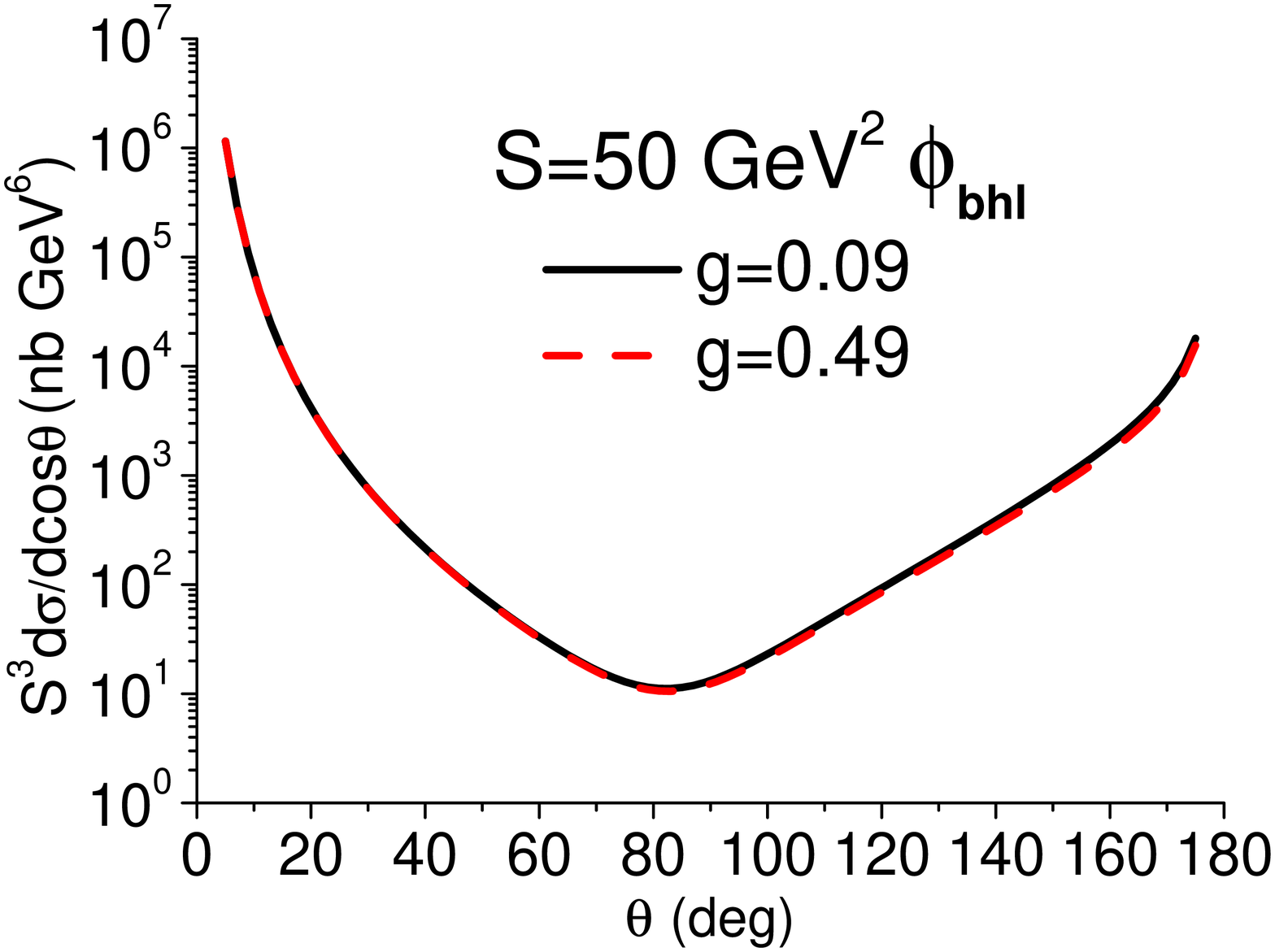}}
\scalebox{0.175}[0.175]{\includegraphics*[36pt,0pt][760pt,590pt]{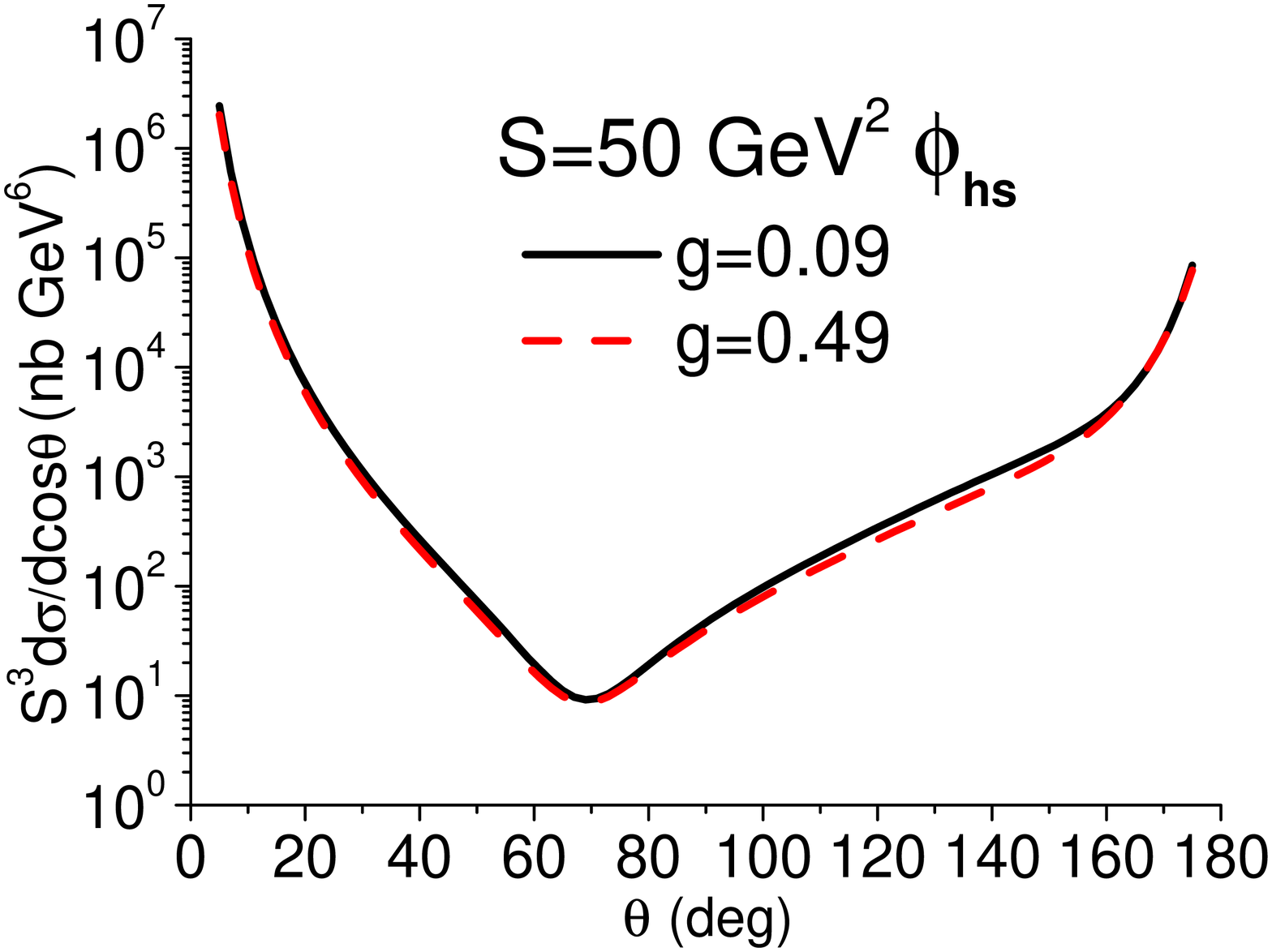}}
\caption{\footnotesize Cross sections at $S=50~\mbox{GeV}^2$.}
\label{50}
\end {figure}

Straightly, we have:
\begin{equation}
\frac{d\sigma}{d\cos \theta}=\frac{1}{32\pi S
(1-v^2)}\Sigma|M|^2=\frac{3}{4}\frac{(2\pi)^3\alpha_e^2(\sqrt{3}f_\pi)^2}{S^3(1-v^2)}\Sigma|M'|^2,
\end{equation}
where $\Sigma|M'|^2=|M'_{LR}+M'_{RR}+(1-1/v)M'_{+R}|^2$, with
$M'$'s calculated from Eq.~(\ref{FT}) with $H'$'s in Table 1 by
including the factor $\alpha_s(\tilde{Q}^2)$ which is not chosen
as a frozen coupling constant \cite{ZM} in this paper. $M_{RR}^2$
contributes dominantly over $M_{LR}^2$ and $M_{+R}^2$. For the
applicability of pQCD, the concerned gluon should be hard, i.e.,
$\tilde{Q}^2$ should be large enough to prevent mis-absorbing soft
contributions. Therefore, we do not fix the QCD running coupling
constant
\begin{equation}
\alpha_s(\tilde{Q}^2)=\frac{4\pi}{b_0\log{\tilde{Q}^2/\Lambda^2}},
\end{equation}
where $b_0=11-2n_f/3$, $n_f$ is the number of active quark
flavors, and $\Lambda\simeq 200$~MeV. We let $\alpha_s$ running if
$\tilde{Q}^2$ is larger than a given $g$ $\mbox{GeV}^2$;
otherwise, we let $\alpha_s=0$, expecting that only pure pQCD
contributions are considered. $\alpha_s(g)$ and factorization
scale $Q_0$ can not be determined {\em a priori} by pQCD theory.
So the choosing of $g$ seems of some kind of arbitrary. However,
the physics should not depend on which cutoff we make, i.e., the
results should not change much if we change $g$ slightly. In
Fig.~\ref{4} we find that in the $\phi_{as}$ case different $g$ do
not affect the result through $10^\circ \to 80^\circ$ and deviate
a little in $80^\circ \to 170^\circ$. In the $\phi_{bhl}$ case the
two curves of different $g$ are even more adjacent because
$\phi_{bhl}$ is just the end-point-suppressed partner of
$\phi_{as}$. $\phi_{cz}$ has bumps near the end points so its
results are very sensitive to $g$ (Fig.~\ref{cz}). It can be
expected that $\phi_{hs}$ will be better than $\phi_{cz}$ at the
cost of suppressing the end-points, just as we show in
Fig.~\ref{4}. With the squared center-of-mass energy $S$
increasing, the gluons have generally more tendency to be hard.
Thus, the parameter $g$ becomes less important. But $\phi_{cz}$ is
still bad even when $S$ reaches $50$~$\mbox{GeV}^2$, therefore we
will not consider it any more as it lacks predictive power. We
calculate the cross sections by using different distribution
amplitudes and varying $g=0.09$~($\alpha_{max}=1.7$), $g=0.49$~
($\alpha_{max}=0.56$) and $S=4,10,20,50$~$\mbox{GeV}^2$
(Figs.~\ref{4}, \ref{10}, \ref{20}, and \ref{50}). Fortunately,
the cross sections are independent to $g$ when the energy scale is
large enough. This leads us to put forward ``work point", a
concept analogical to ``Kondo temperature".
\input epsf

%\section{Discussion}

When the cross sections of one given distribution amplitude are no
longer affected by $g$, we define the present energy scale as the
``work point'' of this distribution amplitude. Different
distribution amplitudes have different work points. The
end-point-suppressed one usually has lower work point. Only above
the work point which means that the contribution from
$\tilde{Q}^2<g$ GeV$^2$ region can be safely neglected and all the
contribution comes mainly from $\tilde{Q}^2$ large region, the
condition for the applicability of pQCD is satisfied and the
calculations of the given process using this distribution
amplitude are possibly reliable. By our work it is clear that the
work points defined by $S$ are, $\phi_{as}$: $\sim
20$~$\mbox{GeV}^2$; $\phi_{bhl}$: $\sim 10$~$\mbox{GeV}^2$;
$\phi_{hs}$: $\sim 20$~$\mbox{GeV}^2$. With the transverse
momentum contributions considered, Sudakov effect \cite{CL} may
reduce the work point lower. If we relax our constraint to a weak
sense, i.e., demanding the cross section difference with different
$g$ should be smaller than some percents, the work point of
$\phi_{bhl}$ and $\phi_{as}$ will be lowered to 4~$\mbox{GeV}^2$
in the small angle region for Fig.~\ref{4}. This is in agreement
with Coriano-Li \cite{CL}.

\begin {figure*}[h]%7
\epsfxsize=7cm\epsfysize=4.8cm\epsfbox{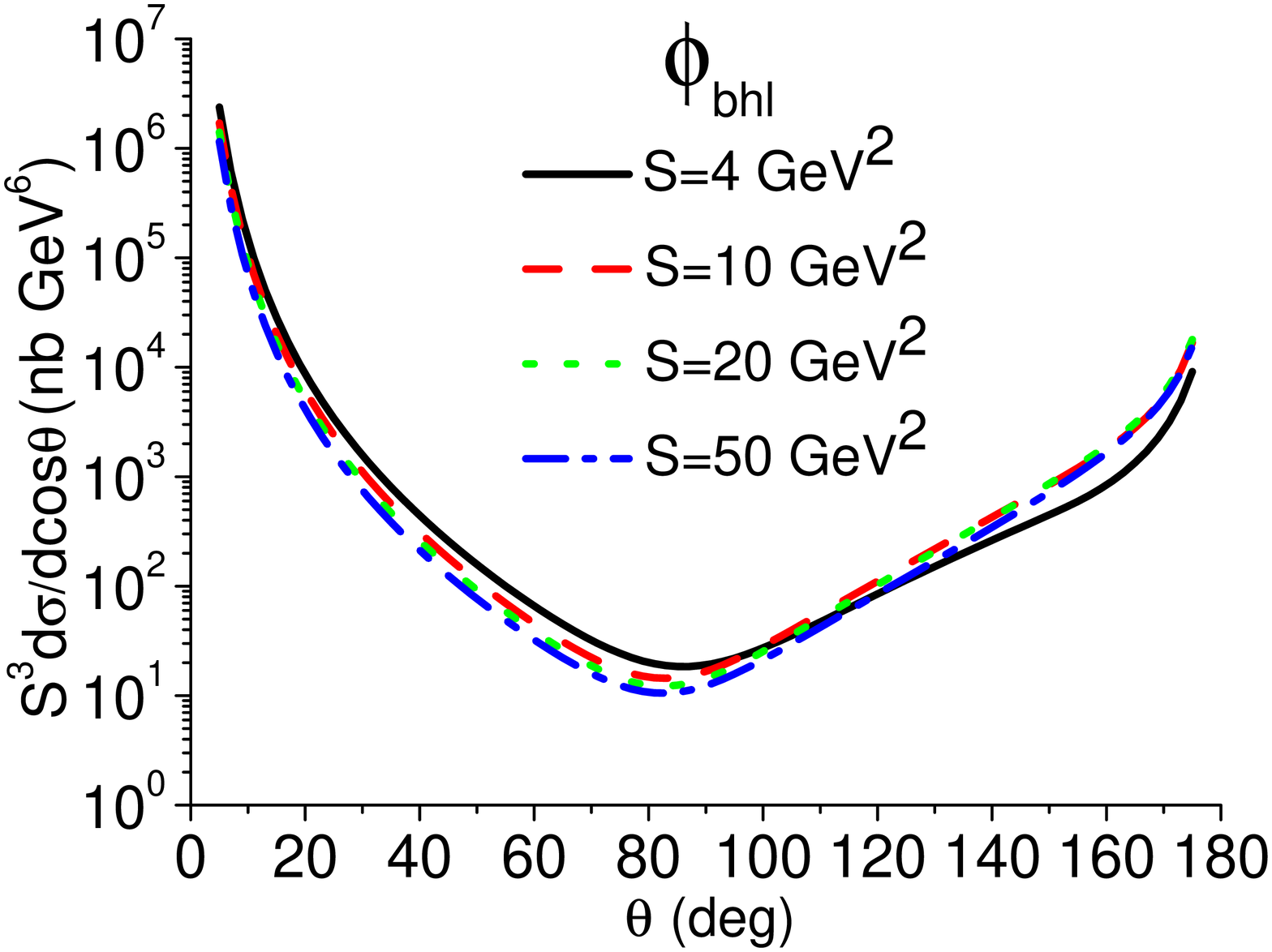}
\epsfxsize=7cm\epsfysize=4.8cm\epsfbox{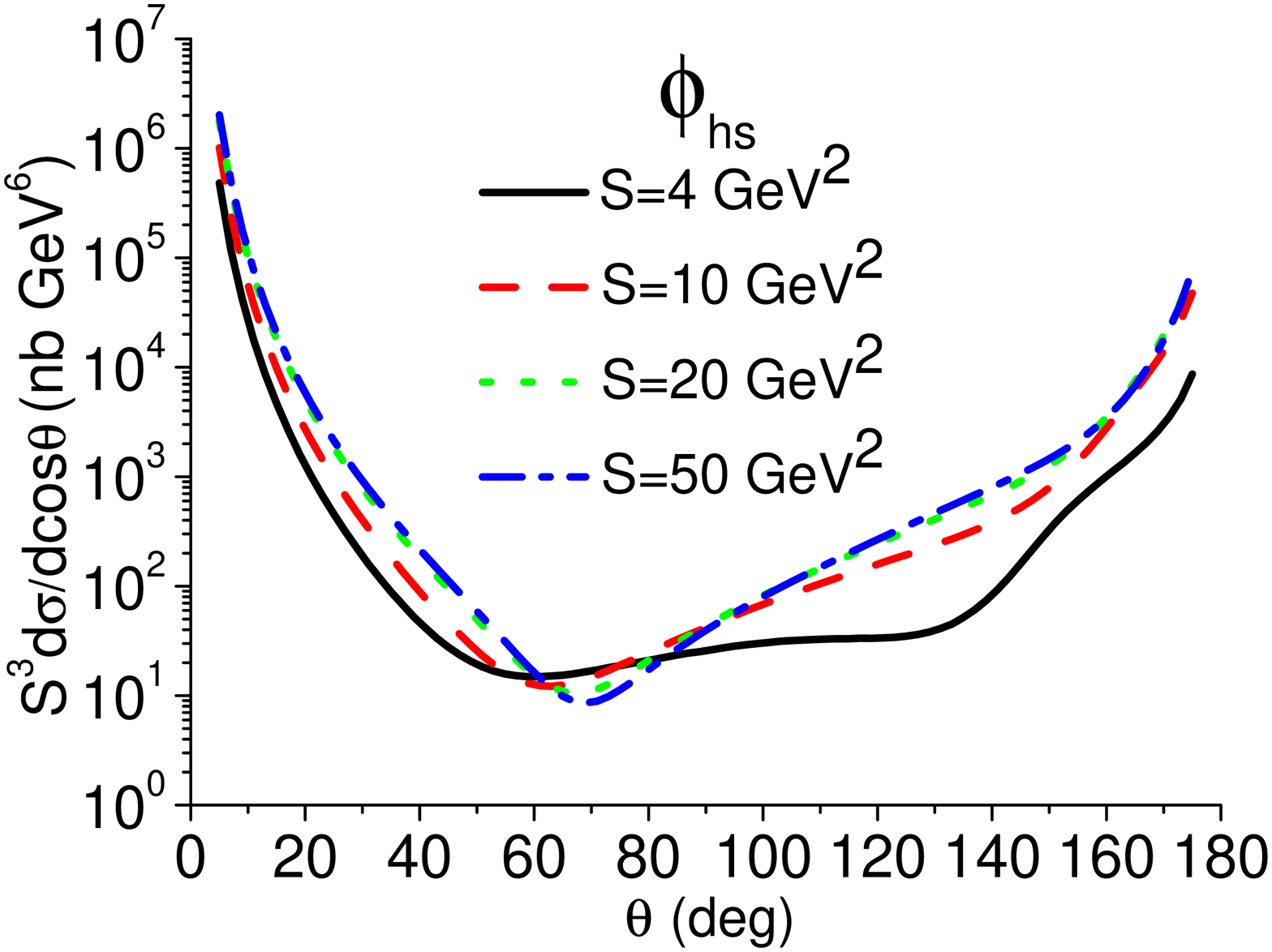}
\caption{\footnotesize Cross sections at different $S$ - scaling
($g=0.49$). } \label{scaling}
\end {figure*}

\begin {figure*}[h]%8
\epsfxsize=7cm\epsfysize=4.8cm\epsfbox{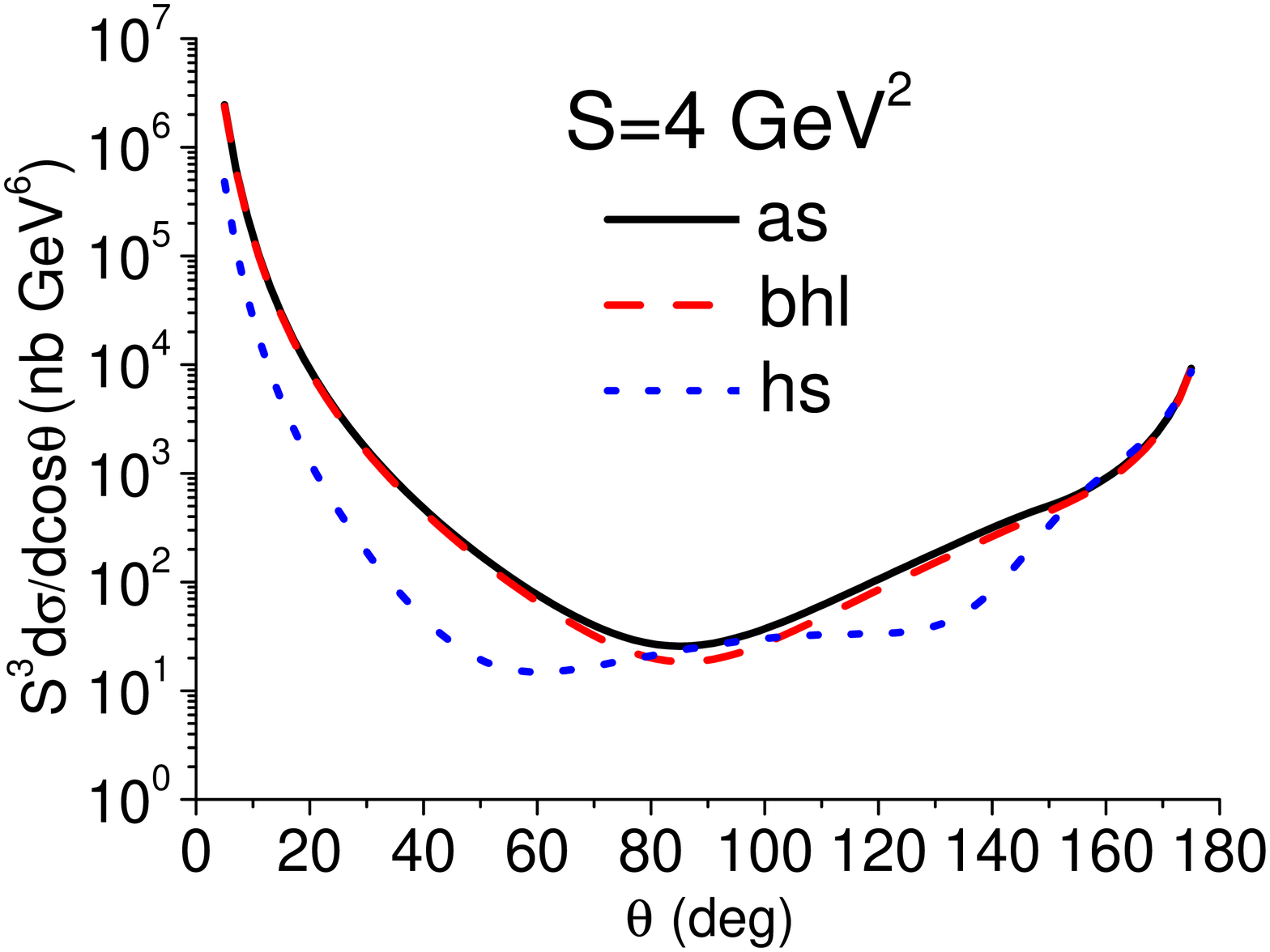}
\epsfxsize=7cm\epsfysize=4.8cm\epsfbox{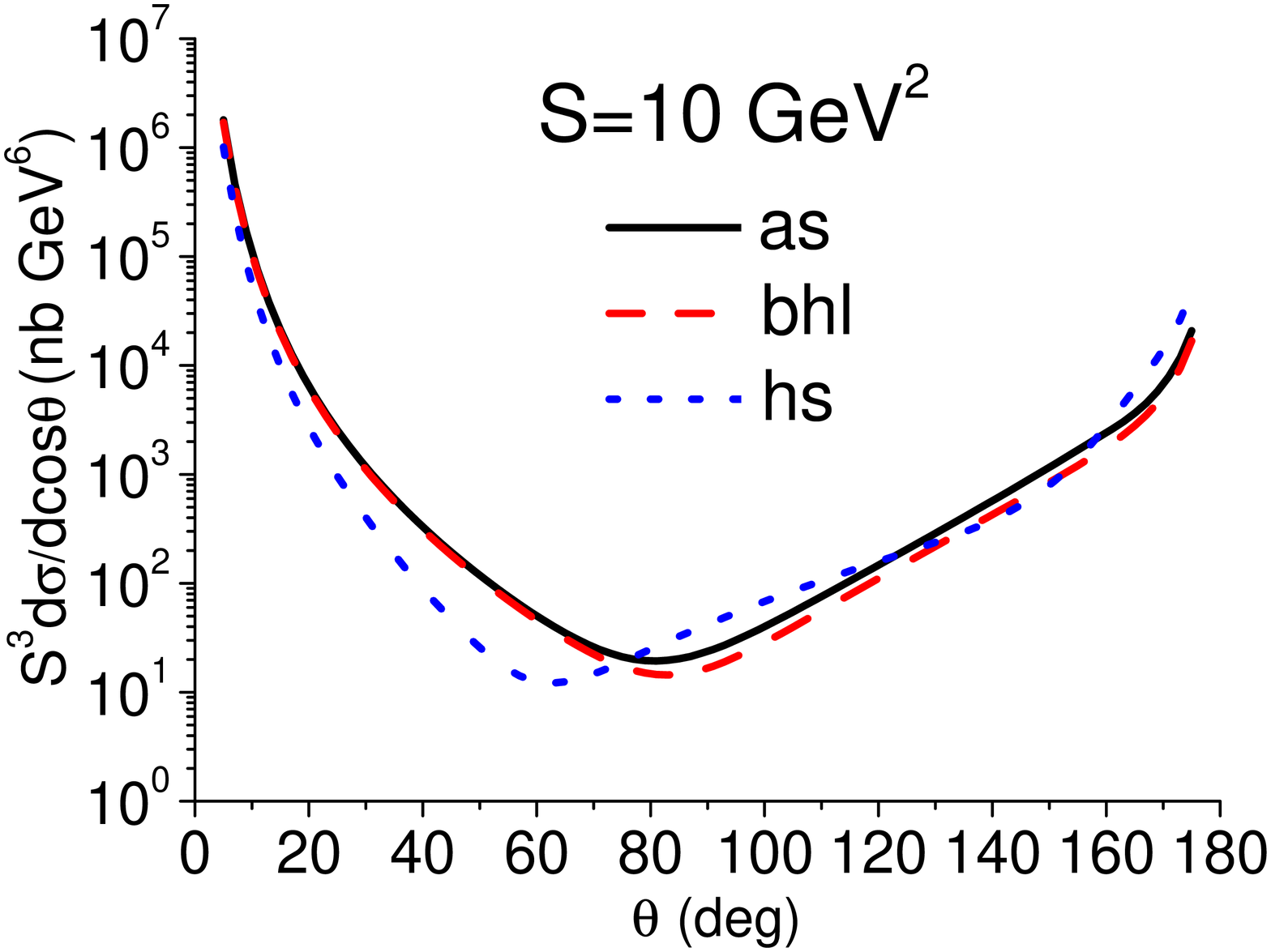}
\epsfxsize=7cm\epsfysize=4.8cm\epsfbox{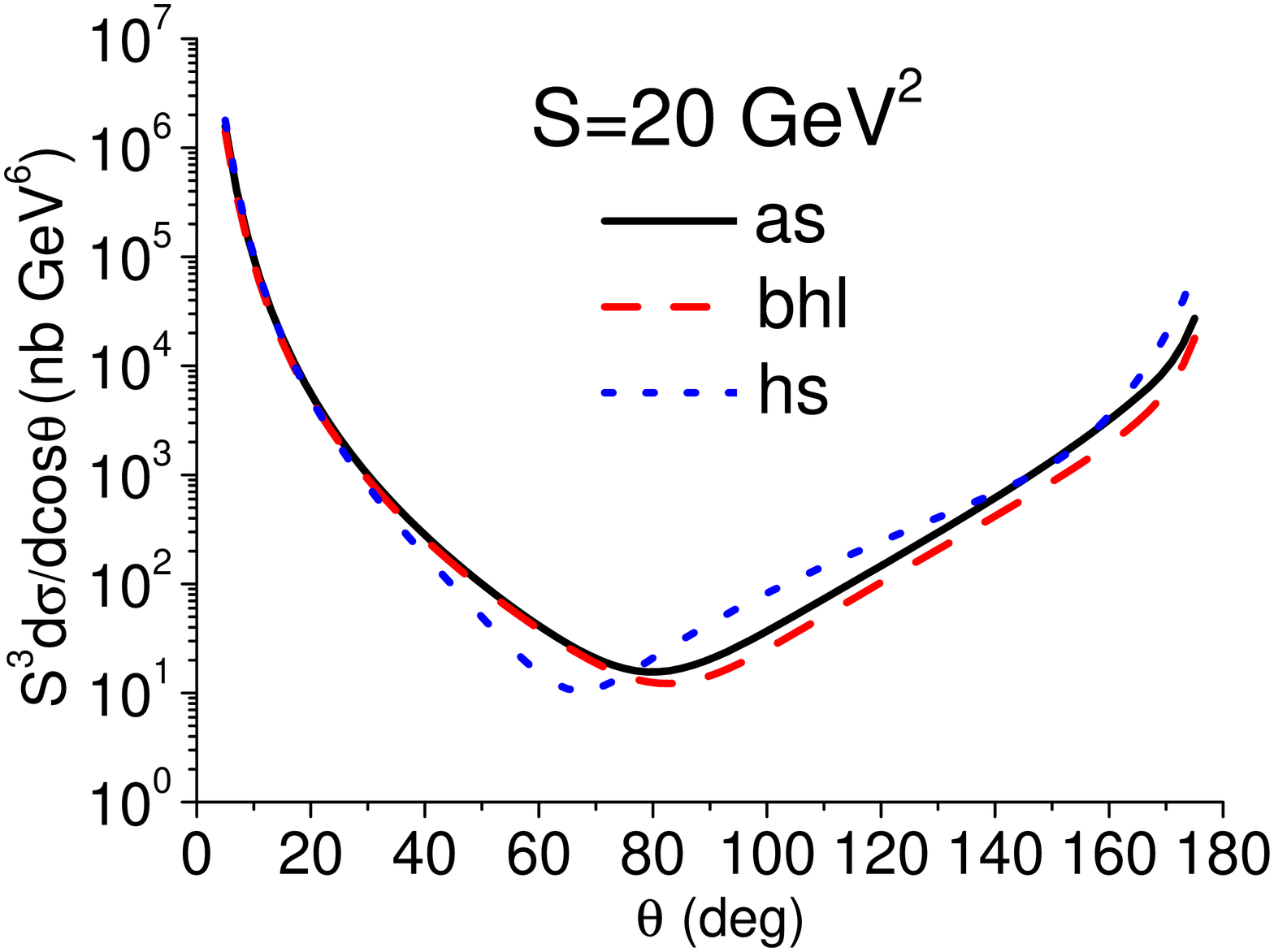}
\epsfxsize=7cm\epsfysize=4.8cm\epsfbox{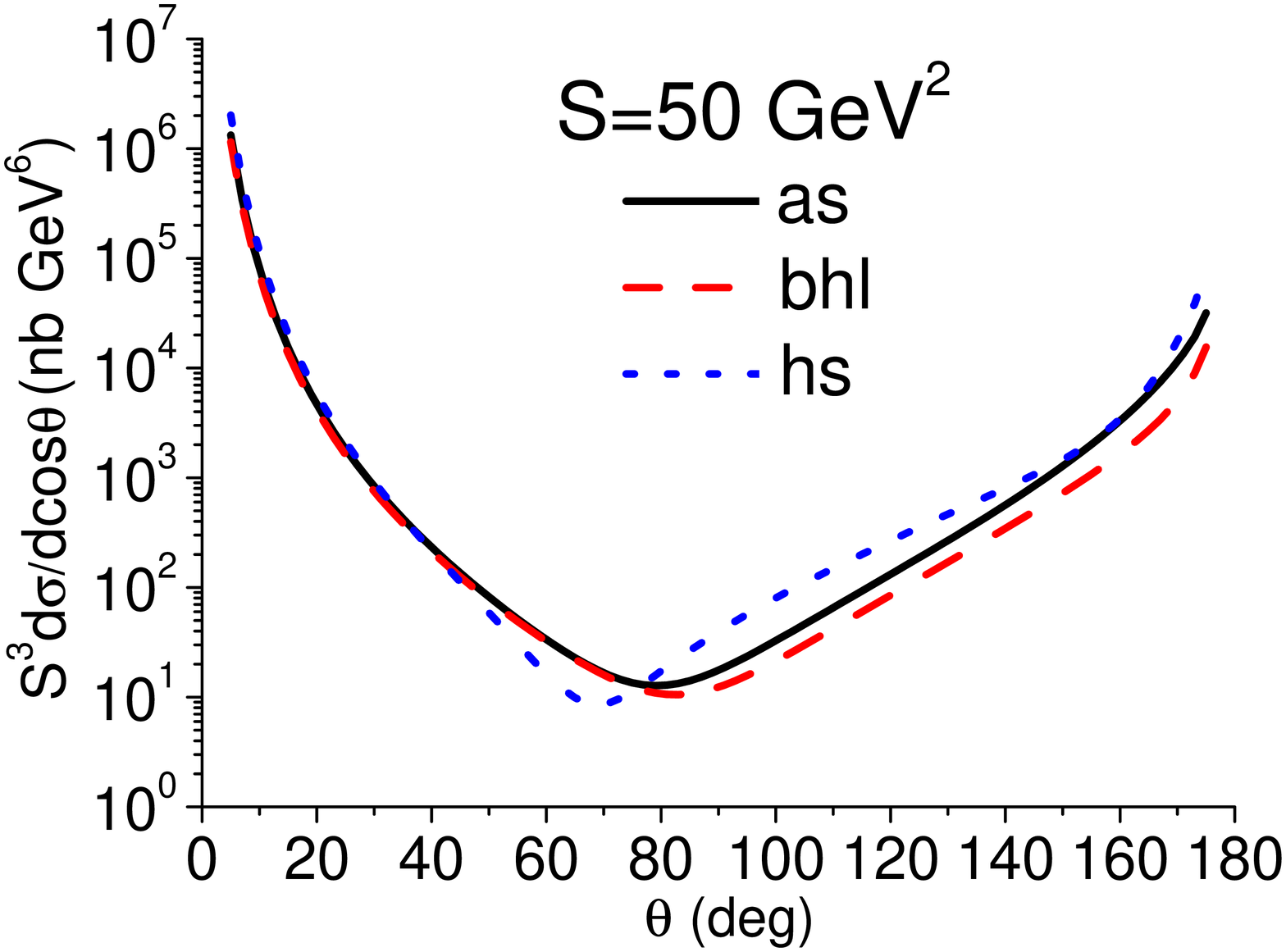}
\caption{\footnotesize Comparison between different distribution
amplitudes ($g=0.49$).} \label{Comparison}
\end {figure*}

As shown in Fig.~\ref{scaling}, we calculate
$S^3\frac{d\sigma}{d\cos \theta}$ with $S$ varying from
4~$\mbox{GeV}^2$ to 50~$\mbox{GeV}^2$. $\frac{d\sigma}{d\cos
\theta}$ obeys approximately the $S^{-3}$ law at fixed angle as
naive pQCD has predicted \cite{BF}. The cutoff of the running
$\alpha_s$ suppresses the value of the cross section with
approximate one order of magnitude, but it renders enough to
survive the scaling law. It is also nontrivial that the cross
sections are almost the same at small angles, especially from
$10^\circ$ to $80^\circ$, for different distribution amplitudes
(Fig.~\ref{Comparison}), as we have no justification to demand
that. On the other hand, they are so apart at large angles, so
that with further experiments we can determine which one should be
chosen in the VCS case. We should pay attention that the cross
sections of $\phi_{hs}$ have a minimum running from $60^\circ$ to
$70^\circ$ while the cross sections of $\phi_{as}$ and
$\phi_{bhl}$ have the same minimum always at $80^\circ$. Summing
up $20^\circ$ to $80^\circ$ to get the ``total" cross section
should be meaningful as predictions, as shown in Table ~\ref{tab
2}.

\renewcommand{\baselinestretch}{1.0}
\begin{table}
\caption{\label{tab 2} Total cross sections [$20^\circ$-
$80^\circ$]}
\begin{tabular}{lccccc} \hline\hline\\
   $S(\mbox{GeV}^2)$&4&10&20&50\\
   \hline\\
   $\sigma$ (nb):$\phi_{bhl}$ & 377 & 16.9 & 1.75 & $9.22\times 10^{-2}$ \\
   $\sigma$ (nb):$\phi_{hs}$ &-- & 6.5 & 1.52 & $1.12\times10^{-1}$ \\
    \hline\hline\\
\end{tabular}
\end{table}
\renewcommand{\baselinestretch}{1.25}

We should state that this is not an exact numerical estimate of
the cross sections, but rather a consistency check, without any
higher-twist contributions or radiative corrections. SELEX
Collaboration at Fermilab \cite{ex} obtained the total forward
cross section of $\pi^{-}e\to\pi^{-}e\gamma$ of 38.8~nb, in
agreement with the theoretical expectation by the chiral
perturbation theory method \cite{U}. Their $S\leq M^2_\rho$, which
is below the work point of any distribution amplitude we have
explored. It illuminates that in the nowadays experimental
available energy region the non-perturbative contribution is
dominant over the pQCD contribution for the pion VCS process.


\begin {figure}[h]%9
\epsfxsize=7cm\epsfysize=4.8cm\epsfbox{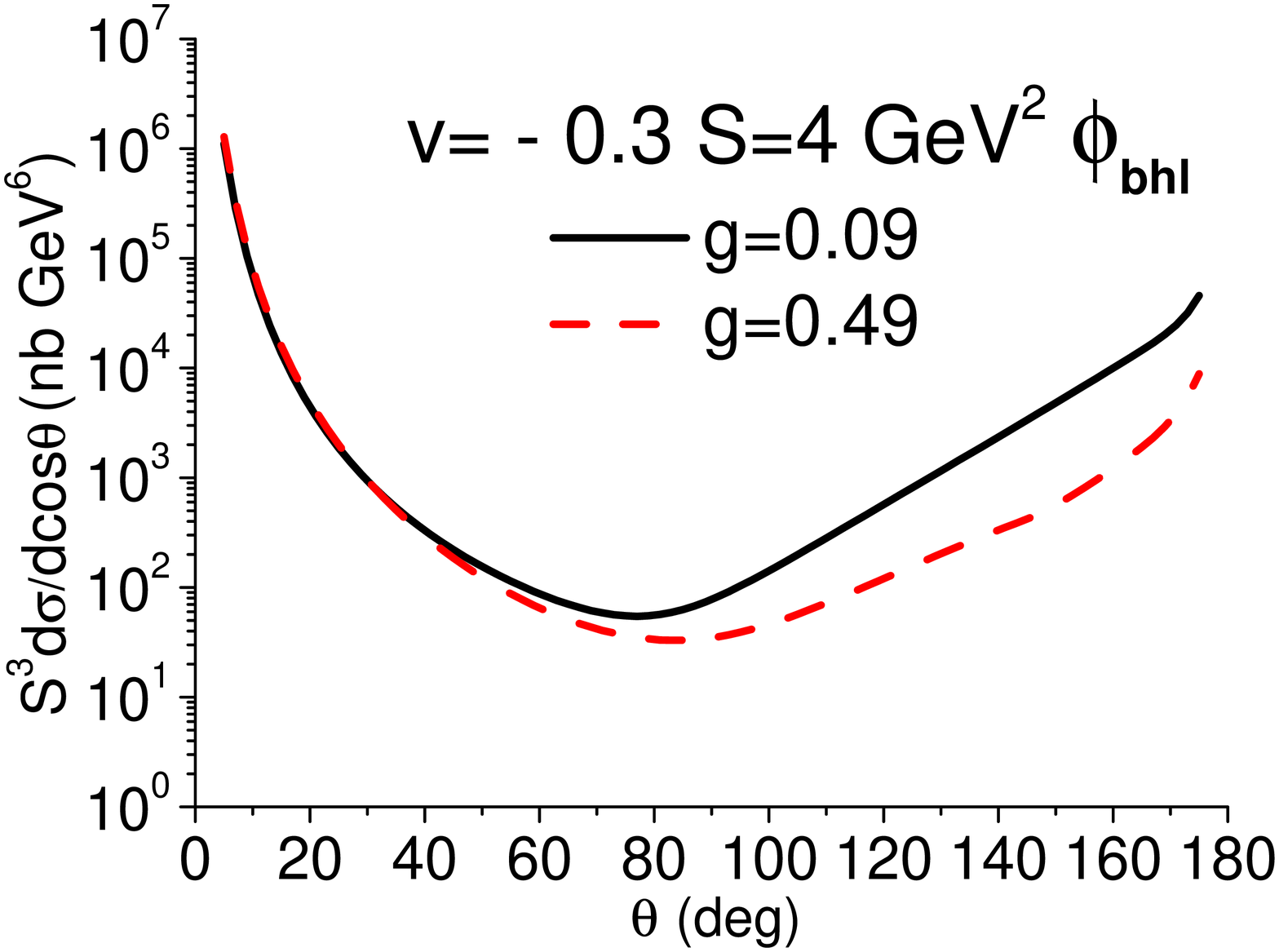}
\epsfxsize=7cm\epsfysize=4.8cm\epsfbox{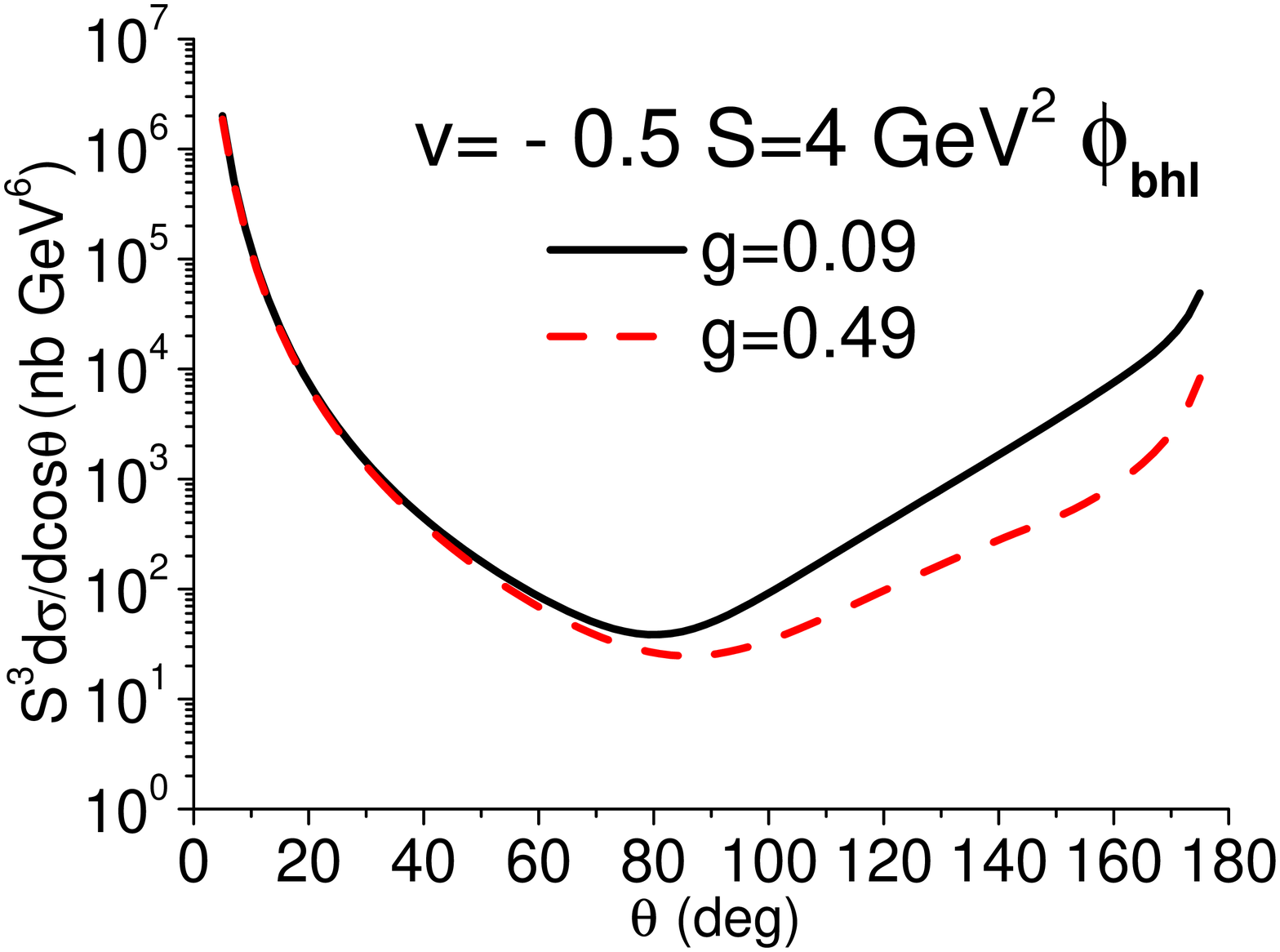}
\caption{\footnotesize Cross sections with different $v$.}
\label{v}
\end {figure}

\begin {figure}[h]%10
\epsfxsize=7cm\epsfysize=4.8cm\epsfbox{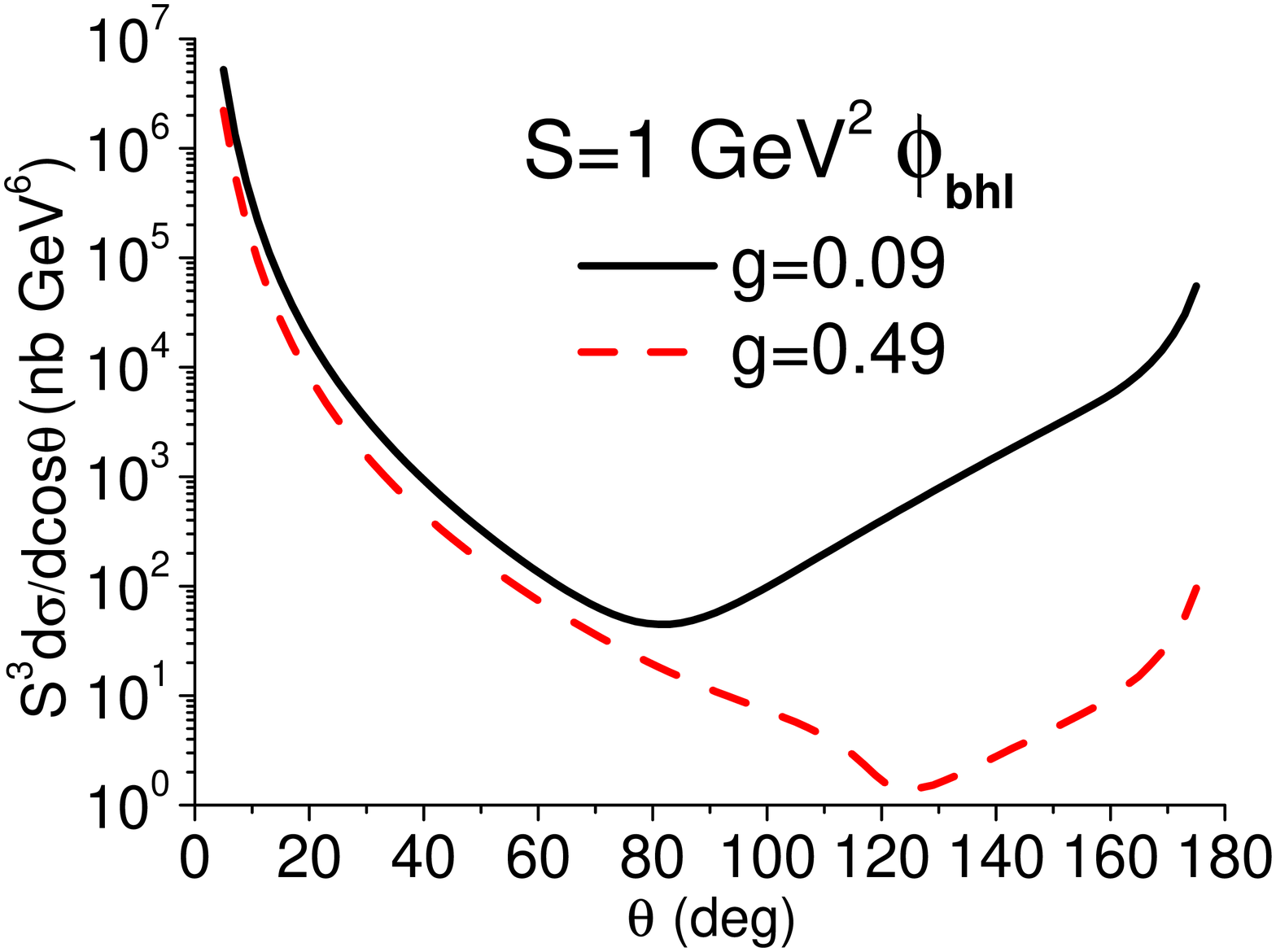}
\epsfxsize=7cm\epsfysize=4.8cm\epsfbox{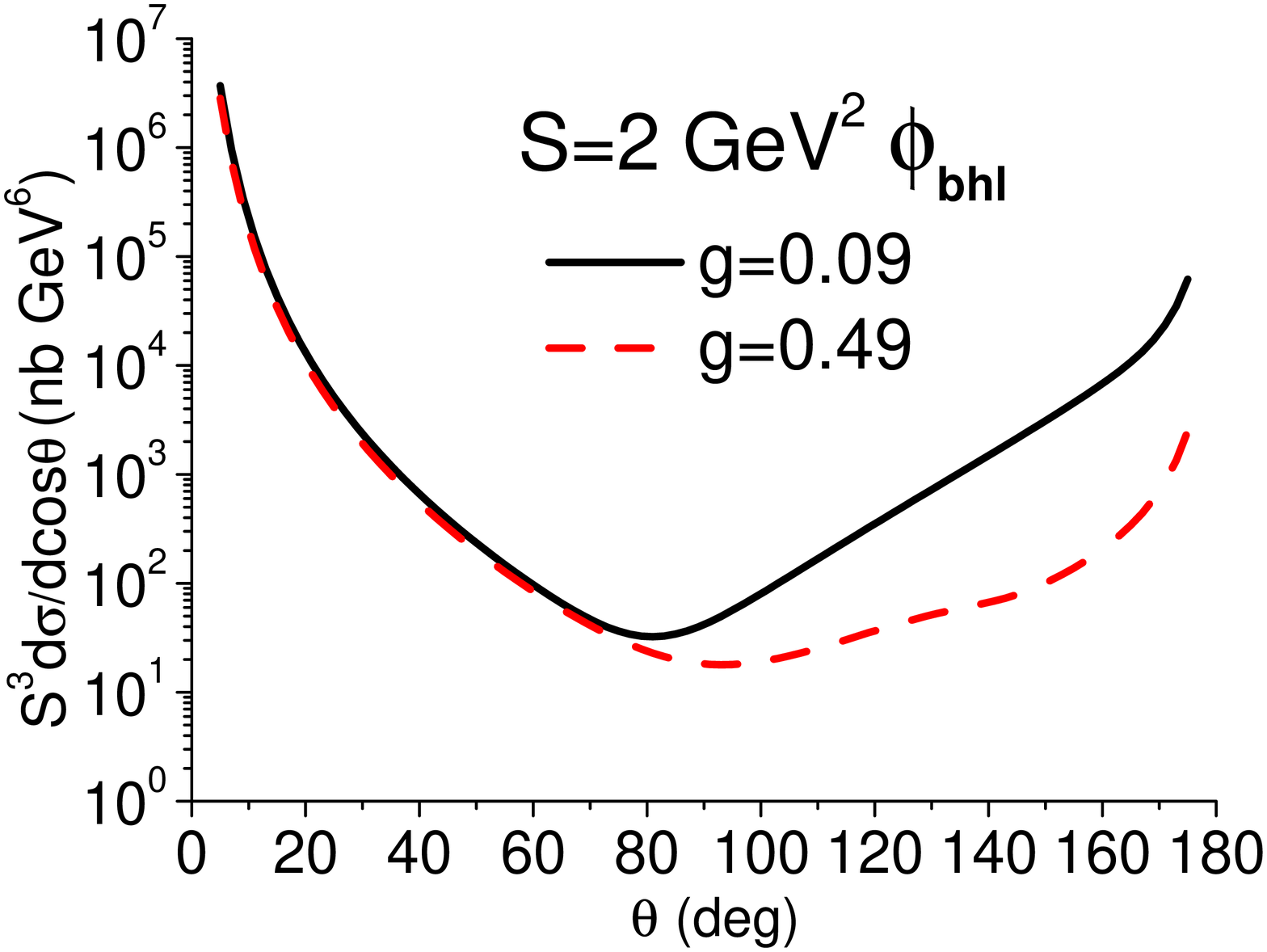}
\caption{\footnotesize Cross sections at lower $S$.} \label{1}
\end {figure}

The dependence on the photon virtuality $v$ is not very sensitive,
as can be seen from Fig.~\ref{v}, where two different values of
$v=-0.3$ and $-0.5$ are examined for $\phi_{bhl}$ at
$S=4~\mbox{GeV}^2$. In other words, the squared virtual photon
4-momentum transfer $q^2=S\, v$ is not a signal energy scale in
virtual Compton scattering. And $S=1~$ and $2~\mbox{GeV}^2$ for
$\phi_{bhl}$ are also examined in Fig.~\ref{1}, from which one may
even arrive at a conclusion for the applicability of pQCD, in a
rather weak sense, for $\phi_{bhl}$ at $S=1\to 2~\mbox{GeV}^2$ in
the region $20^\circ$ to $80^\circ$.

Every distribution amplitude evolves in principle to the
asymptotic form at very high energy scale. Such evolution effects
are small for $\phi_{as}$ and $\phi_{bhl}$, whereas they should be
significant for $\phi_{cz}$ and $\phi_{hs}$. However, there has
been no numerical evaluation on the evolution of $\phi_{cz}$ and
$\phi_{hs}$ in previous studies. It is reasonable to expect a
lower energy scale for the applicability of pQCD for $\phi_{cz}$
and $\phi_{hs}$ than what we claimed, after taking into the
evolution effects.

In summary, we studied explicitly, for the first time, the
applicability of pQCD to the pion virtual Compton scattering. As
in other exclusive processes such as the pion form factor, the
end-point contribution is not negligible in the pQCD calculation
of the pion VCS process, as manifested by the difference between
the cross sections of $\phi_{as}$ with $\phi_{bhl}$ and
$\phi_{cz}$ with $\phi_{hs}$. We also noticed that there exist
middle-region singularities introduced by the QCD running coupling
constant of the exchanged gluons between the two valence partons
of the pion. We introduced a simple technique to evaluate the
contributions from these singularities, so that we can arrive at a
judgement that these contributions will be unharmful to the
applicability of pQCD at certain energy scale, i.e., the work
point to guarantee the safety of pQCD. The work points for
different distribution amplitudes are explored in detail in this
work. For the end-point-suppressed distribution amplitudes such as
$\phi_{bhl}$, we showed that pQCD begins to work at 10
$\mbox{GeV}^2$. If we relax our constraint to a weak sense, the
work point may be as low as $4$ $\mbox{GeV}^2$.

%\section*{Acknowledgments}

{\bf Acknowledgments:} This work is partially supported by
National Natural Science Foundation of China under Grant Numbers
10025523 and 90103007.

%-----------------------------------------------------------------------

% \appendix

% \section{}
% \label{}

% Bibliographic references with the natbib package:
% Parenthetical: \citep{Bai92} produces (Bailyn 1992).
% Textual: \citet{Bai95} produces Bailyn et al. (1995).
% An affix and part of a reference:
%   \citep[e.g.][Ch. 2]{Bar76}
%   produces (e.g. Barnes et al. 1976, Ch. 2).

%\begin{thebibliography}{}

% \bibitem[Names(Year)]{label} or \bibitem[Names(Year)Long names]{label}.
% (\harvarditem{Name}{Year}{label} is also supported.)

%\bibitem[]{}

%\end{thebibliography}


\begin{thebibliography}{99}
%-----------------------------------------------------------------------

\bibitem{BL} G.P. Lepage, S.J. Brodsky, Phys. Lett. B 87 (1979)
359; Phys. Rev. Lett. 43 (1979) 545;

G.R. Farrar, D.R. Jackson, Phys. Rev. Lett. 43 (1979) 246;

A.V. Efremov, A.V. Radyushkin, Phys. Lett. B 94 (1980) 245;

A. Duncan, A.H. Mueller, Phys. Lett. B 90 (1980) 159.
%1

\bibitem{Lep80} G.P. Lepage, S.J. Brodsky, Phys.~Rev.~{D 22} (1980)
2157.%2
\bibitem{BHL81} S.J. Brodsky, T. Huang, G.P. Lepage, in \textit{%
Particles and Fields-2}, Proceedings of the Banff Summer
Institute, Banff, Alberta, 1981, edited by A.Z. Capri and A.N.
Kamal (Plenum, New York,1983), p. 143.%3
\bibitem{IL} N. Isgur, C.H. Llewellyn Smith, Phys. Rev. Lett. 52 (1984) 1080;
Nucl. Phys. B 317 (1989) 526.%4
\bibitem{HS} T. Huang, Q.-X. Shen, Z. Phys. C 50 (1991) 139. %5
\bibitem{LS} H. Li, G. Sterman, Nucl. Phys. B 381 (1992)
129. %6
\bibitem{Yeh}T.-W. Yeh, Phys. Rev. D 65 (2002) 074016.%7
\bibitem{U} C. Unkmeir {\it et al.}, Phys.Rev.C 65 (2002) 015206. %8
\bibitem{CL} C. Coriano, H. Li, Nucl. Phys. B 434 (1995) 535. %9
\bibitem{T} M. Tamazouzt, Phys. Lett. B 211 (1988) 477. %10
\bibitem{MT} E. Maina, R. Torasso, Phys. Lett. B 320 (1994) 337.%11
\bibitem{ZM} D. Zeng, B.-Q. Ma, Phys. Lett. B 542 (2002) 55. \\
It should be ``$1-s^2(y+\bar{v}x)+2\bar{v}xys^4$" in
the bracket of the numerator of the hard part $h$ in Table 1 of this paper.%12
\bibitem{CZ}
V.L. Chernyak, A.R. Zhinitsky, Nucl. Phys. B 201 (1982) 492.%13
\bibitem{Huang94} T. Huang, B.-Q. Ma, Q.-X. Shen, Phys. Rev. D {49} (1994)
1490.%14
\bibitem{BF} S.J. Brodsky, G.R. Farrar, Phys. Rev. Lett. 31 (1973)
1153.%15
\bibitem{ex} SELEX Collaboration, A. Ocherashvili {\it et al.}, Phys. Rev. C 66
(2002) 034613. %16
\end{thebibliography}
\end{document}